\documentclass{amsart}
\usepackage{graphicx}
\usepackage{boxedminipage}
\usepackage{bm}
\theoremstyle{definition}
\newtheorem{defn}{Definition}

\newtheorem{rem}{Remark}

\numberwithin{equation}{section} \numberwithin{ex}{section}
\numberwithin{exr}{section} \numberwithin{defn}{section}
\numberwithin{rem}{section}\numberwithin{mybox}{section}
\numberwithin{figure}{section}

\newcommand{\Real}{\mathbb R}

\begin{document}

\title[]{Relativistic anelasticity}%
\author{Marcelo Epstein}%
\address{Department of Mechanical and Manufacturing Engineering \\
University of Calgary \\
Calgary, Alberta T2N 1N4, Canada}%
\email{mepstein@ucalgary.ca}%

\author{David A Burton}%
\address{Department of Physics\\
University of Lancaster \\
Lancaster, LA1 4YB, UK}%
\email{d.burton@lancaster.ac.uk}%

\author{Robin Tucker}
\address{Department of Physics\\
University of Lancaster \\
Lancaster, LA1 4YB, UK}%
\email{r.tucker@lancaster.ac.uk}%

\thanks{}%
\subjclass{}%
\keywords{}%

\begin{abstract}
\vspace{0.4cm}{\small A formulation of Continuum Mechanics within
the context of General Relativity is presented that allows for the
incorporation of certain types of anelastic material behaviour,
such as viscoelasticity and plasticity. The approach is based on
the concept of a four-dimensional body-time complex structured as a
principal bundle. The anelastic response is regarded as the result
of a continuous distribution of inhomogeneities, whose evolution
is dictated by a suggested relativistic version of the Eshelby
tensor. The role played by various groups is emphasized throughout
the presentation and illustrated by means of the example of an
anelastic fluid. }
\end{abstract}
\maketitle

\section{Introduction}
\newcommand{\mcM}{\ensuremath{{\mathcal M}}}
\newcommand{\mcN}{\ensuremath{{\mathcal N}}}
\newcommand{\mcE}{\ensuremath{{\mathcal E}}}
\newcommand{\mcU}{\ensuremath{{\mathcal U}}}
\newcommand{\mcV}{\ensuremath{{\mathcal V}}}
\newcommand{\mcD}{\ensuremath{{\mathcal D}}}
\newcommand{\mcB}{\ensuremath{{\mathcal B}}}
\newcommand{\mcF}{\ensuremath{{\mathcal F}}}
\newcommand{\mcJ}{\ensuremath{{\mathcal J}}}
\newcommand{\mcS}{\ensuremath{{\mathcal S}}}
\newcommand{\mcT}{\ensuremath{{\mathcal T}}}
\newcommand{\PD}{\ensuremath{{\partial}}}
\newcommand{\Bg}{\ensuremath{{\mathbf g}}}
\newcommand{\BG}{\ensuremath{{\mathbf G}}}
\newcommand{\Bomega}{\ensuremath{{\bm{\omega}}}}
\newcommand{\Bmu}{\ensuremath{{\bm{\mu}}}}
\newcommand{\Balpha}{\ensuremath{{\bm{\alpha}}}}
\newcommand{\BY}{\ensuremath{{\mathbf Y}}}
\newcommand{\BT}{\ensuremath{{\mathbf T}}}
\newcommand{\Bu}{\ensuremath{{\mathbf u}}}
\newcommand{\Bt}{\ensuremath{{\mathbf t}}}
\newcommand{\Lie}[1]{\ensuremath{{\mathcal L}_{#1}}}
\newcommand{\SecD}{\ensuremath{{\mcJ_{\BG,\lambda}\sigma}}}
\newcommand{\SecDGvar}{\ensuremath{{\mcJ_{\BG_\varepsilon,\lambda}\sigma}}}
\newcommand{\SecDlambdavar}{\ensuremath{{\mcJ_{\BG,\lambda_\varepsilon}\sigma}}}
\newcommand{\DiffSecD}[2]{\ensuremath{{\mcJ_{{#2}_*\BG,{#1}^*\lambda}{#1}^*{\sigma}}}}
\newcommand{\DiffSecDabbrev}[1]{\ensuremath{{\mcJ^{#1}_{\BG,\lambda}\sigma}}}
There are many situations in physics where a knowledge of the
thermo-mechanical bulk properties of matter is essential for an
understanding of a wide variety of natural phenomena. Among these
properties the response of matter to internal and external forces
is often a dominant feature that determines its static or dynamic
behaviour. The laws of classical (i.e., non-relativistic)
Continuum Mechanics have been developed to address such phenomena
and constitutive theory is now a mature branch of material
science. The formulation of traditional non-relativistic Continuum
Mechanics is firmly rooted in the framework of Newtonian dynamics
and time dependent transformations that preserve the Euclidean
structure of a three dimensional space are deeply ingrained in its
insistence on the {\it principle of material objectivity}
\cite{NLFT} and compatibility with thermodynamics. Furthermore,
the evolution and past history of material deformation depend on a
universal Newtonian time  and are therefore incompatible with
notions of observer dependence in a relativistic spacetime. Whilst
such discrepancies may be irrelevant for some problems, they
should not in principle be ignored. They cannot be ignored for
problems in which relative material speeds comparable with that of
light are relevant or for problems involving non-Newtonian
accelerations or large gravitational tidal stresses. Such
situations can arise in numerous problems in astrophysics and
space science.  Modelling the behaviour of rapidly rotating
neutron stars offers a challenging avenue for exploring new
material phenomena as do situations involving the interaction of
matter  with intense electromagnetic or gravitational radiation.
The constitutive properties of materials in such situations are
far from clear since they depend on particular reformulations of
the laws of non-relativistic Continuum Mechanics that must be
compatible with the local Lorentz structure of spacetime needed to
maintain the causal propagation of physical effects within all
media.

A number of relativistic reformulations exist. Based on pioneering
efforts by Rayner \cite{Ray}, Carter and Quintana \cite{Car},
Maugin \cite{Mau}, Kijowski \cite{Kij}, and others, most
reformulations model the spacetime histories of a three-dimensional
material medium by a congruence of time-like worldlines in
spacetime $\mathcal M$ and identify a material body $\mathcal B$
with a class of three-dimensional spacelike surfaces transverse to
such a congruence. It is assumed that equations can be prescribed
to determine a map $f: {\mathcal M} \rightarrow {\mathcal B} $
such that in any coordinates $X^A$ for some domain of $\mathcal B$
the components $f^A$ of $f$ satisfy $d\,f^1\wedge
d\,f^2\wedge\,d\,f^3\ne 0$. Using the spacetime metric $\bf g$,
the $3-$form $d\,f^1\wedge d\,f^2\wedge\,d\,f^3$ is mapped to a
future pointing time-like normalized $4-$velocity field, the
integral curves of which model the world lines of material
elements of $\mathcal B$ in $\mathcal M$. The equations for $f$
can be derived from a variational action principle or (for
dissipative phenomena) by setting to zero the divergence of a
suitable stress-energy tensor on $\mathcal M$. The applications of
this formalism are largely dictated by how the action or
stress-energy tensor are constructed in terms of $f$ and the
spacetime metric $\bf g$. Since, in general, spacetime has no
particular Killing isometries the role of the Euclidean structure
in non-relativistic Continuum Mechanics is at best a guiding
principle in the modelling process and a new principle of
objectivity must be found within covariances of the spacetime
formulation.

The dissipative behaviour of many materials can be modelled within
the context of classical (i.e., non-relativistic) Continuum
Mechanics by means of an elastic prototype, namely, a fixed
elastic material ``point''. The most striking example of this
situation is provided by the theory of ideal elasto-plasticity,
whereby the process of plastification does not alter the elastic
constants of the material, as evidenced by subjecting the material
to an unloading process. What does change during plastification
is, therefore, the way in which the elastic prototype is implanted
into the body as time goes on. A similar interpretation can be
given to processes of biological bulk growth, such as those
observed in bone and muscle. Here, the anelasticity reflects the
fact that, as time goes on, more (or less) material of the same
type is squashed into the body, the properties of the new material
being identical to those of the existing one. As in the case of
plasticity, the result is, in general, the development of
deformations and of residual stresses. We may say that in these
phenomena there are two different types of kinematics at play: the
ordinary kinematics of embeddings of the body manifold into space
and the material kinematics resulting from the time variation of
the implants. These two kinematical paradigms are quite
independent from each other. Correspondingly, the forces behind
them are of diverse natures. As much as ordinary (Cauchy or Piola)
stresses can be seen as the causes of the evolution of the spatial
kinematics, the forces reponsible for the material evolution are
the so-called configurational stresses, pioneered by Eshelby as
far back as 1951 in his classical paper \cite{Esh} dealing with
the force acting on an elastic singularity. Assuming we start from
an originally perfectly homogeneous stress-free body and we
subject it to some loading process, if the material is of the type
just described, the different points may undergo a material
evolution (plastification, say), each point to a different extent
than the others in its neighbourhood. The result (after unloading,
for example) is that, although the body is still {\it uniform} in
the sense that it is made of the same material at all of its
points, it can no longer be considered homogeneous: it has an
irreversible continuous distribution of inhomogeneities in its
midst. We may say that this type of inhomogeneity and the
previously described type of anelasticity are the material and
spatial counterparts of each other.

In any relativistic framework it becomes necessary  to accommodate
material features that have a natural interpretation in a
non-relativistic limit in terms of classical hysteretic
dissipation, plasticity, material inhomogeneity, aging and more
general anelastic properties that involve memory effects and rate
dependent responses to the environment. In this paper we begin an
exploration of these issues in terms of a new formulation of
relativistic Continuum Mechanics based on the concept of a
four-dimensional body-time complex $\mathcal N$ rather than the
three-dimensional body manifold $\mathcal B$. The essential idea in
our relativistic formulation is to use a rank $4$ map $\kappa:
{\mathcal N} \rightarrow {\mathcal M}$ to describe the physics of
a deformable body. Embeddings of lower rank have been used to
define relativistic point particles, strings and membranes
\cite{rwt1}, \cite{rwt2}, \cite{rwt3}, \cite{rwt4}, \cite{rwt5},
\cite{rwt6} and for higher dimensional ${\mathcal N}$ and
$\mathcal M$ (and reparameterisation invariant actions), which
form the cornerstone of current efforts to model the basic
interactions. It is our intention to base a general relativistic
theory of constitutive properties of materials on the map $\kappa$
that offers a simpler approach for a relativistic description of
anelasticity.

\section{The material body-time manifold} \label{sec:body-time}

In classical Continuum Mechanics, the notion of a material body
$\mathcal B$ as a three-dimensional differentiable manifold can be
said to emerge from the following considerations. All possible
spatial manifestations (or configurations) of an identifiable
entity happen to be connected open sets in $\Real^3$ with the
property: there exists a unique and smooth correspondence between
the points of each and every pair of configurations. We call the
corresponding equivalence classes {\it material points}. Since the
correspondence has been assumed to be smooth with respect to
spatial positions, it follows that each configuration can be seen
as an atlas of an underlying differentiable manifold $\mathcal B$,
called the material body. It is common to assume that there exists
an atlas consisting of just one chart. The configurations can now
be regarded as embeddings of the material body in $\Real^3$. This
fastidious heuristic ``derivation'' of the notion of a classical
material body will serve the purpose of justifying the following
relativistic counterpart.

In General Relativity, the object under consideration manifests
itself always as a world tube $\mathcal T$ consisting of a
congruence of world lines in a relativistic four-dimensional
space-time manifold $\mathcal M$. The assertion that this is a
congruence of world lines and not just a world tube is a
reflection of the fact that, just as in the classical counterpart,
we postulate that particles are distinguishable from each other.
Moreover, on each of these lines there exists a
unique-up-to-translation parametrization such that:
\begin{equation}
{\bf g}({\bf u}, {\bf u})=-1,
\end{equation}
where $\bf g$ is the metric of $\mathcal M$ and $\bf u$ is the
future-pointing timelike unit vector field tangent to the world
line. In each of these tubes, therefore, there exists a physically
meaningful (local) action of the group of translations
of $\Real$ and, for the sake of
brevity, we denote this group also by $\Real$.
The manifold of integral curves of $\bf u$ is the quotient ${\mathcal
B}={\mathcal T}/{\Real}$ and we are naturally led to the
following definition:
\begin{defn}{\rm A {\it material body-time} (or {\it body-time
complex}) is a principal fibre bundle $({\mathcal N},{\mathcal
B},\pi,\Real,\Real)$.
The total manifold $\mathcal N$ alone will be referred to as the
material body-time manifold. The base manifold $\mathcal B$ will
be assumed to be an ordinary material body, namely, a trivial
three-dimensional manifold. The projection map:
\begin{equation}
\pi: \mathcal N \longrightarrow \mathcal B,
\end{equation}
assigns to each body-time point the corresponding body point. The
principal nature of the bundle reflects itself in that the typical
fibre $\Real$ and the structural group coincide.}
\end{defn}

A material tangent vector ${\bf V} \in T\mathcal N$ is said to be
{\it vertical} if
\begin{equation}
\pi_*({\bf V})={\bf 0}.
\end{equation}
\newcommand{\Fibreg}{{\bf g}^{\Real}}
We endow the fibre bundle $({\mathcal N},{\mathcal
B},\pi,\Real,\Real)$ with the fibre metric $\Fibreg = -d S
\otimes d S$ where $S$ is a fibre coordinate and define on
$\mathcal N$ a vertical vector field ${\bf U}=\frac{\PD}{\PD S}$ satisfying
\begin{equation}
\Fibreg({\bf U},{\bf U}) = -1.
\end{equation}

\section{Histories}

A history is the relativistic counterpart of a classical
configuration. As discussed above, therefore, a material history
can be identified with a congruence of world lines. More formally,
we state the following:
\begin{defn} {\rm A {\it history} is an embedding
\begin{equation}
\kappa: \mathcal N \longrightarrow \mathcal M,
\end{equation}
such that:
\begin{equation} \label{eq:compatibility}
{\bf g}(\kappa_*({\bf U}),\kappa_*({\bf U}))=-1.
\end{equation}
}
\end{defn}
Since ${\bf U}=\frac{\PD}{\PD S}$ Equation (\ref{eq:compatibility})
identifies $S$ with the proper
time of each integral curve of $\kappa_*({\bf U})$. 
The tangent map
\begin{equation}
\kappa_* : T\mathcal N \longrightarrow T\mathcal M,
\end{equation}
appearing in Equation (\ref{eq:compatibility}) is called the {\it
history gradient} of $\kappa$. At each point $n \in \mathcal N$ it
is represented by a tensor{\footnote{ Since the vector spaces
connected by this tensor are associated with different manifolds
it is often referred to as a {\it two-point} tensor} ${\bf
F}=\kappa_*:T_n\mathcal N \longrightarrow T_{\kappa(n)}\mathcal
M$.

\begin{rem} {\rm For simplicity, we are assuming that a history is an
embedding of the whole of $\mathcal N$. In reality, one may have
to consider histories that consist of embeddings of some time-wise
portion of $\mathcal N$.}
\end{rem}
We will often refer to Equation (\ref{eq:compatibility}) as the
{\it time consistency condition} of our embeddings. It represents
a fundamental constraint in the dynamical description, as we shall
soon see.

\begin{rem} It would also have been possible not to require the
satisfaction of the time-consistency condition. In other words,
one can derive an alternative theory whereby the structural group
of $\mathcal N$ is the group of all (orientation-preserving)
diffeomorphisms of the real line. It is not entirely clear at this
point whether both formulations are equivalent up to a
re-parametrization of solutions. This may very well be the case
under some constitutive restrictions (some of which are discussed
in later sections), but it appears that in the most general
situation the consistency condition (and its associated Lagrange
multiplier) play a fundamental role in the theory to be presented.
\end{rem}

It may prove convenient to exhibit the local coordinate versions
of some of the basic quantities and equations introduced so far.
All coordinate systems in $\mathcal N$ will be assumed to be
consistent with a trivialization of the fibre bundle. The four
coordinates, therefore, will be split into three body-coordinates
$X^\Lambda,\;\Lambda=1,2,3,$ and the proper time-coordinate $S$. For
compactness, we will also often adopt the notation $X^A,
\;A=1,...,4,$ with $X^4=S$. In a spacetime domain, on the other
hand, we will denote the coordinates by lower-case letters:
$x^a,\;a=1,...,4$. Thus, a history $\kappa$ is given in
coordinates in terms of four smooth functions:
\begin{equation}
x^a={\kappa}^a(X^A)={\kappa}^a(X^\Lambda,S),
\end{equation}
and the corresponding history gradient is represented by the field
of matrices:
\begin{equation}
F^a_A=\frac{\partial {\kappa}^a}{\partial X^A}.
\end{equation}
This matrix field is of maximal rank (4), since we have assumed
the histories to be embeddings. We will also assume it has a
positive determinant. The image $\kappa_* {\bf U}$ of the
canonical unit vector field $\bf U=\frac{\partial}{\partial S}$
provides us with the four-velocity field $\bf u$ on $\mathcal M$,
given in components by:
\begin{equation}
u^a=F^a_A U^A.
\end{equation}
With ${\mathcal M}$ time-orientable $\bf u$ will be declared
future-pointing for $S>0$.  Since the only non-vanishing component
of $\bf U$ is $U^4=1$, condition (\ref{eq:compatibility}) becomes
in components :
\begin{equation} \label{eq:constraint1}
g_{ab}F^a_4F^b_4=-1.
\end{equation}

\section{Elasticity}

In classical Continuum Mechanics, a first-grade (or simple)
material point without memory effects is characterized by
constitutive laws (for the stress, the heat flux, the internal
energy and the entropy) whose independent variables are the
present local values of the deformation gradient, the temperature
and the temperature gradient. A material is said to be {\it
elastic} if the stress tensor depends on the deformation gradient
alone. Thus, in the classical setting, it is possible to remain
within the realm of a purely mechanical theory, whose field
equations are the conservation of mass and the balance of linear
(and angular) momentum and whose constitutive response is just an
elastic law for the stress tensor. In a similar vein, one may
avoid a complete and explicit thermodynamic treatment even in a
theory of viscous flow, by allowing time-derivatives of the
deformation gradient to enter the picture, but ignoring the
temperature and its gradient. If an elastic material is subjected
to an {\it internal constraint}, such as incompressibility or
inextensibility along a material direction or any other
restriction imposed on its deformation gradient, the stress is not
completely determined by the deformation gradient, since the
forces (stresses) necessary to maintain the constraint are not
related to the deformation. For example, if a material die is
inextensible in the direction of one of its sides, one may apply
an arbitrary surface traction in the inextensible direction
without producing any deformation! The total stress is then
assumed to be resolvable into the sum of two parts: one determined
by the deformation (according to some elastic law) and another one
given by a Lagrange multiplier applied to an expression dictated
by the constraint equation. The standard way to obtain this
expression (see \cite{NLFT}, Section 30) is based on the
assumption that the second part does no work in any motion
satisfying the constraint. In other words, the constraint forces
do not expend mechanical power on the admissible virtual
velocities. In the case of holonomic constraints, this criterion
leads to the gradient of the constraint equation. More precisely,
let $\phi({\bf F})=0$ be the given constraint equation, where $\bf
F$ is now the \emph{classical} deformation gradient. Then the term to be
added to the determinate part of the stress is of the form:
\begin{equation} \label{eq:multiplier}
\lambda \; \frac{\partial \phi}{\partial {\bf F}} {\bf F}^T,
\end{equation}
$\lambda$ being the Lagrange multiplier.

In General Relativity the thermomechanical response of a material
is assumed to be encoded in the symmetric stress-energy-momentum
tensor $\bf t$. In trying to emulate the classical theory of
elastic simple materials, one can assume that the value of $\bf t$
at a point $m$ of $\mathcal M$ occupied by a point $n \in \mathcal
N$ depends exclusively on $n$, on the local value (at $m$) of the
metric $\bf g$ and on the local value (at $n$) of the history
gradient $\bf F$. This a-priori exclusion of thermal effects may
need some further discussion. Be that as it may, it is important
to realize that in our kinematical picture {\it all materials are
to be considered as internally constrained}. The reason for this
somewhat surprising assertion lies in our time-consistency
condition (\ref{eq:compatibility}), which is a (universal)
restriction on all possible history gradients. Following the
paradigm of classical Continuum Mechanics as given in Equation
(\ref{eq:multiplier}), we obtain the following general form of the
stress-energy-momentum tensor of a relativistic elastic material:
\begin{equation} \label{eq:elastic}
{\bf t}={\bf f}(n,{\bf g},{\bf F})+\lambda(n)\;{\bf u} \otimes
{\bf u},
\end{equation}
where, as before, ${\bf u}={\bf F U}$, and where $\bf f$ is an
arbitrary {\it symmetric} tensor-valued function of the arguments
shown. The particular form of the term affected by the Lagrange
multiplier $\lambda$ in (\ref{eq:elastic}) arises when applying
the prescription (\ref{eq:multiplier}) to the constraint equation
(\ref{eq:constraint1}), namely with:
\begin{equation} \label{eq:constraint2}
\phi=g_{ab}F^a_AF^b_BU^AU^B+1.
\end{equation}
Note that Equation (\ref{eq:constraint1}) is imposed (i.e. $\phi$ is
set to zero) after calculating (\ref{eq:multiplier}).

An elastic constitutive law is said to be {\it hyperelastic} if
$\bf f$ derives from a scalar potential $W=W(n,{\bf g},{\bf F})$
according to the formula:
\begin{equation} \label{eq:hyperelastic1}
{\bf f}=W {\bf g}^{-1}+\frac{\partial W}{\partial {\bf F}}{\bf
F}^T,
\end{equation}
or, in components:
\begin{equation}
f^{ab}=W\;g^{ab}+g^{ad}\frac{\partial W}{\partial F^d_A}\;F^b_A.
\end{equation}
The quantity $W$ corresponds to the classical {\it strain energy
per unit spatial volume}. In general relativity, therefore, we
interpret it as the elastic potential {\it per unit space-time
volume} as induced by the metric, namely by the 4-form in
$\mathcal M$ with component $\sqrt{-\det[g_{ab}]}$ in a coordinate
basis. An alternative approach to determine the general form of
the hyperelastic stress-energy-momentum tensor is via a
constrained variational principle, as shown in the Appendix.

\begin{rem} In classical Continuum Mechanics it is customary to
introduce a so-called {\it reference configuration} and to induce
the Cartesian volume therein onto the body manifold. In that case,
it makes sense to define a strain energy density per unit volume
in the body (but this volume depends of the reference
configuration chosen through the determinant of the gradient of
the change of reference). Alternatively, a volume form is
introduced in the body representing the mass density, in which
case one can introduce a strain-energy per unit mass. By defining
various pulled-back stress tensors or tensor densities, the
expression (\ref{eq:hyperelastic1}) is simplified to the extent
that it loses its ``spherical'' part, retaining only the term with
the partial derivative of the potential (see \cite{NLFT}, Section
82). Similar arguments can be used in general relativity, the most
widespread being that of the baryon number, assumed to be
conserved. We will not pursue such arguments at this point,
preferring to stick to the space-time volume form, which is
supposed to be available regardless of any other considerations.
\end{rem}

\section{Reduced elastic constitutive law}

There are different physical and mathematical criteria that can be
invoked to reduce the general form of the elastic constitutive law
(\ref{eq:elastic}). The first and most important of these is the
{\it principle of covariance}. We will demand that the functional
expressions of the constitutive laws be covariant (that is,
tensorial) under any change of frame, whether holonomic or not, in
spacetime $\mathcal M$. To find the corresponding reduction in the
form of the constitutive law, we commence by noting that under a
change of frame given by a matrix with entries $S^i_a$, the frame
components of the history gradient change according to:
\begin{equation}
{\hat F}^i_A=F^a_A S^i_a,
\end{equation}
where a hat indicates the components in the new frame. Similarly,
the frame components of the metric vary according to:
\begin{equation}
{\hat g}_{ij}=(S^{-1})^a_i(S^{-1})^b_j\;g_{ab}.
\end{equation}
Suppose at first that $\psi$ is a scalar constitutive quantity
prescribed by a law on $\mathcal N$ of the form
\begin{equation}
\psi={\bar \psi}(X^A,g_{ab},F^a_A),
\end{equation}
where $F^a_A$ and $g_{ab}$ are evaluated at $X^A$ and
${\kappa}^a(X^A)$, respectively. We suppress the dependence on
material tensors or constants on $\mathcal N$ since these remain
inert under changes of frame on $\mathcal M$. We demand that in
the new frame the value of the function $\bar \psi$ remain
unchanged, namely\footnote{%
A jet bundle-theoretic derivation of
(\ref{eq:component_equivariance}) based on an equivariance
property under diffeomorphisms is given in the Appendix.}:
\begin{equation} \label{eq:component_equivariance}
\psi={\bar \psi}(X^A,g_{ij},F^i_A)={\bar
\psi}(X^A,(S^{-1})^a_i(S^{-1})^b_j\;g_{ab},F^a_AS^i_a).
\end{equation}
This is an identity to be satisfied for all choices of the
non-singular matrix $\{S^i_a\}$. We may, therefore, choose its
entries to coincide with those of ${\bf F}^{-1}$, thus obtaining:
\begin{equation}
\psi={\tilde \psi}(X^A,F^i_AF^j_B\;g_{ij}),
\end{equation}
where $\tilde \psi$ is an arbitrary function. In conclusion, the
principle of covariance requires that:
\begin{equation} \label{eq:scalar}
\psi={\tilde \psi}({\bf X}, {\bf C}),
\end{equation}
where $\bf C = {\bf F}^T{\bf gF}$ is the pull-back (by the history
$\kappa$) of the metric $\bf g$ to the body-time
manifold.\footnote{It is interesting to note that this result
coincides with the classical one (derived from the classical
principle objectivity or frame-indifference), by a subtle
balancing act between the new variables (namely, the presence of a
non-trivial metric) and the changes of frame allowed (namely,
arbitrary changes, rather than just orthonormal).}

\begin{rem} {\rm {\bf An interesting identity:} A direct
consequence of the reduced form of a scalar constitutive law is
the existence of an identity satisfied by its various partial
derivatives. It is not difficult to show, by simply using the
chain rule of differentiation and effecting a few algebraic
operations, that Equation (\ref{eq:scalar}) implies:
\begin{equation}
{\bf F} \left(\frac{\partial {\tilde \psi}}{\partial {\bf
F}}\right)^T= 2\left(\frac{\partial {\tilde \psi}}{\partial {\bf
g}}\right){\bf g},
\end{equation}
or, for the sake of clarity, in components:
\begin{equation}
\label{eq:derivative_component_equivariance}
F^i_I\frac{\partial {\tilde \psi}}{\partial F^j_I}=2\frac{\partial
{\tilde \psi}}{\partial g_{ia}}g_{ja}.
\end{equation}
It is quite possible that (perhaps modulo some topological
condition) the reasoning can be reversed: the satisfaction of this
identity would then imply the specific dependence of the
constitutive law on $\bf C$, rather than on any other combination
of $\bf F$ and $\bf g$.}
\end{rem}

If we now repeat the previous reasoning for a space-time
tensor-valued function (such as the stress-energy-momentum tensor
{\bf t}), the only difference arises from the fact that the
components of $\bf t$ are affected by the change of frame
according to the rules for second-order tensors. The final reduced
form of the constitutive law (\ref{eq:elastic}) is:
\begin{equation} \label{eq:reduced}
{\bf t}= {\bf F} \;[\;{\tilde {\bf f}}({\bf X}, {\bf C})+\lambda
\;{\bf U} \otimes {\bf U}\;]\;\;{\bf F}^T\;,
\end{equation}
where ${\tilde {\bf f}}$ is a purely body-time tensor which, by
analogy, may be called the (determinate part of the) Kirchhoff
stress-energy-momentum.\footnote{Note that this tensor differs
from the so-called {\it second Piola-Kirchhoff stress} by the
absence of the determinant of the deformation gradient. In fact,
the second Piola-Kirchhoff stress depends on the choice of
reference configuration and can be more properly understood as a
tensor density.}

If the constitutive law happens to be hyperelastic, in accordance
with Equation (\ref{eq:hyperelastic1}) and with the chain rule of
differentiation, the determinate part of the Kirchhoff
stress-energy-momentum tensor is given by:
\begin{equation} \label{eq:hyper2}
{\tilde {\bf f}}=W{\bf C}^{-1}+2\frac{\partial W}{\partial {\bf
C}}.
\end{equation}

It is worth noting that the formal reduction of constitutive laws
just effected is independent of the fibred nature of the body-time
manifold.

Further restrictions on the constitutive law may arise from a
number of causes, some of which we briefly discuss at this point,
leaving a more detailed discussion for later. Since the
time-consistency condition can be expressed as $C_{44}=-1$, it
follows that, in any given system of coordinates, there is no need
to include a dependence on $C_{44}$. This observation can be used,
in combination with other criteria, to further reduce the
constitutive law. Another restriction may arise from physical
reasoning whereby, if dissipation is to be ruled out, the vector
$\bf u$ should be an eigenvector of $\bf t$. Finally, further
reductions can be obtained by invoking additional material
symmetries (such as isotropy) that the material may have. Before
returning to these matters, we proceed to the formulation of
explicit dynamic equations.

\section{The dynamic equations}

A fully fledged dynamical problem calls for the solution of
Einstein's equation:
\begin{equation} \label{eq:einstein}
{\bf Ein}={\bf t},
\end{equation}
where ${\bf Ein}$ is the Einstein tensor associated with the
metric $\bf g$. The right-hand side of this equation incorporates
the constitutive law, itself a function of $\bf g$ through its
dependence on $\bf C$, according to Equation (\ref{eq:reduced}).
But, since the Einstein tensor is divergence-free, we need to
impose the following integrability condition:
\begin{equation} \label{eq:dynamic}
{\bf \nabla \cdot t}=0,
\end{equation}
or, in coordinates:
\begin{equation} \label{eq:dinamic1}
t^{ab}_{\;\;;b}=0,
\end{equation}
where a semicolon indicates covariant differentiation with respect
to the (torsion-free) metric connection. The left-hand side of
this equation is ultimately expressible, via the constitutive law
(\ref{eq:reduced}), in terms of the kinematical variables.
Consequently, Equations (\ref{eq:einstein}) and (\ref{eq:dynamic})
constitute a total of fourteen partial differential equations
which, together with the universal constraint
(\ref{eq:compatibility}), provide the field equations to be solved
for the fifteen unknowns $g_{ab}$, $\kappa^a$ and $\lambda$. When
the presence of the elastic body is likely to affect the
gravitational background only slightly, one may use a perturbation
technique whereby, as a first step, the metric $\bf g$ is kept at
a fixed value and only the dynamic equations (\ref{eq:dinamic1})
are solved (always in conjunction with the constitutive laws and
the constraint) for $\kappa^a$ and $\lambda$ as functions of
$X^A$. This first step, often called the fixed-background problem,
is usually considered sufficient for non-cosmological
applications, for which a very slight violation of Einstein's
equation is certainly tolerable.

There are many useful manipulations afforded by sundry pull-backs
and projections of the field equations. The most common one, and
one that can readily be interpreted physically, consists of
resolving Equation (\ref{eq:dynamic}) into a scalar component on
the local four-velocity vector $\bf u$ and the remaining
projection on its normal three-dimensional hyperplane. Our
objective has been to show that at the level of generality
maintained so far a consistent and complete theory emerges and
that the kinematical variables, including the Lagrange multiplier,
are ultimately obtainable as a solution of the field equations
supplemented with the equation of constraint and the constitutive
law. An example of some importance to evidence the physical
meaning of the Lagrange multiplier is that of dust, that is, a
continuous collection of material particles without any mutual
elastic interaction. In this extreme case, we naturally set the
constitutive function ${\tilde {\bf f}}$ in (\ref{eq:reduced}) to
zero and obtain the following residual form of the constitutive
law:
\begin{equation} \label{eq:dust}
{\bf t} = \lambda {\bf u} \otimes {\bf u},
\end{equation}
which can be recognized as the standard stress-energy-momentum
tensor of a dust, provided one identifies $\lambda$ with the
mass-energy density. For the sake of the exercise, if we now
enforce the dynamic equation (\ref{eq:dynamic}) and take its inner
product with $\bf u$, namely,
\begin{equation} \label{eq:dust1}
{\bf g} \left({\bf u}, {\bf \nabla \cdot} (\lambda {\bf u} \otimes
{\bf u})\right)=0,
\end{equation}
and if we take account of the constraint
(\ref{eq:compatibility}),we readily obtain the scalar equation:
\begin{equation} \label{eq:dust2}
{\bf \nabla \cdot}(\lambda {\bf u})=0,
\end{equation}
which, when subtracted from (\ref{eq:dust1}), yields the geodesic
equation:
\begin{equation} \label{eq:dust3}
{\nabla}_{\bf u}\;{\bf u} = {\bf 0}.
\end{equation}

\section{A drastic reduction of the constitutive law}
\label{sec:drastic}

Given a space-time (scalar-, vector- or tensor-valued)
constitutive function $\Psi(n, {\bf g}, {\bf F})$, a {\it
constitutive symmetry} at the point ${ n} \in \mathcal N$ is an
automorphism $\bf H$ of the tangent space $T_n\mathcal N$ such
that:
\begin{equation} \label{eq:symmetry}
\Psi(n, {\bf g}, {\bf F})=\Psi(n, {\bf g}, {\bf FH}),
\end{equation}
for all history gradients $\bf F$ at $n$. The physical meaning of
a symmetry is that the particular body-time point $n$ is
indifferent in its constitutive response to the pre-application of
a transformation $\bf H$ of its neighbourhood. Notice that this
transformation applies only to the deformation gradient, and not
to the various material tensors which are the repository of the
material properties. It is not difficult to show that all the
symmetries of a constitutive law at $n$ form a group ${\mathcal
H}_n$, which is called the {\it symmetry group} of $\Psi$ at $n$.
In the particular case of the constitutive law (\ref{eq:reduced}),
a symmetry $\bf H$ must satisfy the following identity:
\begin{equation} \label{eq:symmetry1}
{\tilde {\bf f}}(n, {\bf C})+\lambda \;{\bf U} \otimes {\bf
U}={\bf H} \;{\tilde {\bf f}}(n, {\bf H}^T{\bf C H})\;{\bf
H}^T+\lambda \;({\bf HU}) \otimes ({\bf HU}).
\end{equation}
The first conclusion that imposes itself, by virtue of the
independence of the term governed by the Lagrange multiplier from
the determinate term, is that:
\begin{equation} \label{eq:symmetry2}
{\bf HU}=\pm{\bf U},
\end{equation}
or, in other words, that $\bf U$ must be an eigenvector of $\bf H$
with a unitary eigenvalue. This conclusion is consistent with the
fact that other transformations are already ruled out by the
time-consistency condition (\ref{eq:compatibility}). Excluding,
moreover, transformations involving time reversal, we may say that
the eigenvalue is $+1$. In a body-time chart (consistent, as
always, with the fibred structure of $\mathcal N$), a symmetry
$\bf H$ with components $H^A_B$ must have the following general
matrix expression:
\begin{equation} \label{eq:symmetry3}
{\bf H} = \left[
\begin{matrix}
{\bf K} & {\bf 0}  \\
{\bf h}^T & 1
\end{matrix}
\right].
\end{equation}
Here, $\bf K$ is a non-singular $3 \times 3$-matrix and ${\bf
h}^T$ is a $3$-row. Within the group of matrices having this
particular form the symmetry group must lie as a subgroup and, in
principle, it could be as small as the trivial group (consisting
of just the identity transformation).

We now define a particular symmetry group of (\ref{eq:reduced}),
which we will denote by ${\mathcal H}^e$. It consists of all
automorphisms $\bf H$ (having, of course, $\bf U$ as an
eigenvector with unit eigenvalue) that preserve the bundle
projection, namely:
\begin{equation} \label{eq:symmetry4}
\pi_*({\bf HV})=\pi_*{\bf V},\;\;\;\;\forall\,\,{\bf V} \in
T_n{\mathcal N}.
\end{equation}
In a coordinate representation, a typical element of this group
looks as follows:
\begin{equation} \label{eq:symmetry5}
{\bf H}= \left[
\begin{matrix}
{\bf I} & {\bf 0}  \\
{\bf h}^T & 1
\end{matrix}
\right] ,
\end{equation}
with the arbitrary $\bf K$ having been replaced by the unit
matrix. Although a perfectly valid formulation of elastic
materials (according to our definition) can be pursued without
restricting in any way the symmetry group (beyond the general
condition (\ref{eq:symmetry2})), most authors seem to implicitly
assume that every elastic material must have a symmetry group
large enough to contain ${\mathcal H}^e$. This restriction does
not seem to arise from any mathematical consideration, but rather
from a putatively physical reasoning that, by identifying the lack
of symmetry of this type with the sensitivity of the material to a
relative time-shift between neighbouring particles, detects a
source of energy dissipation. Be that as it may, it will become
presently obvious that, in order to relate our formulation to such
well-established works as those of Carter and Quintana \cite{Car}
or Beig and Schmidt \cite{Sch}, the condition
\begin{equation} \label{eq:symmetry6}
{\mathcal H}_n\supseteq {\mathcal H}^e
\end{equation}
will have to be adopted.

We will now investigate the rather severe restrictions imposed by
condition (\ref{eq:symmetry6}) on the elastic generic constitutive
law (\ref{eq:reduced}). The term involving the Lagrange multiplier
needs no further reduction. We are, therefore, left with the
identity:
\begin{equation} \label{eq:symmetry7}
{\tilde {\bf f}}(n, {\bf C})={\bf H} \;{\tilde {\bf f}}(n, {\bf
H}^T{\bf C H})\;{\bf H}^T,
\end{equation}
to be satisfied by all non-singular symmetric tensors $\bf C$
(with $C_{44}=-1$) and all ${\bf H} \in {\mathcal H}^e$. As a
preliminary calculation, we will consider the simpler question of
finding the restrictions that would apply to a {\it scalar}
constitutive law at a point $n \in \mathcal N$, namely, we intend
to find the most general form of a scalar-valued function
$\psi=\psi({\bf C})$ satisfying the identity:
\begin{equation} \label{eq:symmetry8}
\psi({\bf C}) = \psi ({\bf H}^T {\bf C H}),
\end{equation}
for all ${\bf H} \in {\mathcal H}^e$. This is certainly a simpler
problem than that posed by Equation (\ref{eq:symmetry7}). We will
carry out the proof of our formula in a local frame and, at the
end of the process, we will provide an invariant representation of
the result. Let, then, the matrix expression of $\bf C$ in a local
body-time frame (always consistent with the fibration) be given
by:
\begin{equation} \label{eq:symmetry9}
{\bf C} = \left[
\begin{matrix}
{\bf A} & {\bf b} \\
{\bf b}^T & -1
\end{matrix}
\right] ,
\end{equation}
where condition (\ref{eq:compatibility})) has been used in the
form $C_{44}=-1$. The $3 \times 3$ matrix $\bf A$ is symmetric and
positive definite. We now evaluate the matrix product ${\bf
H}^T{\bf C H}$ directly from Equations (\ref{eq:symmetry5}) and
(\ref{eq:symmetry9}) as:
\begin{equation} \label{eq:symmetry10}
{\bf H}^T {\bf C H}= \left[
\begin{matrix}
{\bf A}+{\bf bh}^T+{\bf h b}^T- {\bf hh}^T & {\bf b}- {\bf h} \\
{\bf b}^T - {\bf h}^T & -1
\end{matrix}
\right].
\end{equation}
Since $\bf h$ is arbitrary, for any given $\bf C$ we may certainly
choose:
\begin{equation} \label{eq:symmetry11}
{\bf h} = {\bf b}
\end{equation}
whence it follows that the function $\psi$ must depend on its
arguments through the following peculiar combination:
\begin{equation} \label{eq:symmetry12}
\psi({\bf C})=\psi\left( \left[
\begin{matrix}
{\bf A} +{\bf bb}^T & 0 \\
0 & -1
\end{matrix}
\right] \right).
\end{equation}
One may wonder whether further reduction might still arise from
other choices of $\bf h$. Nevertheless, it is a straightforward
matter to verify that an arbitrary function of the form
(\ref{eq:symmetry12}) satisfies the initial identity, so that
further reduction is not implied by the identity. One may now
question the correctness of this formula on the grounds that it
may not be invariant under changes of (fibre-consistent) body-time
frames. One way to dispel this fear is to check directly that
under a change of frame the form of the formula is preserved. A
more illuminating alternative is obtained by recalling that, given
a non-singular symmetric rank 2 covariant tensor $\bf Z$, we can
canonically define a rank 2 contravariant counterpart ${\bf
Z}^{-1}$, and vice-versa. The matrix representations of these
entities in mutually dual bases are (as suggested by the notation)
inverses of each other. It is now a straightforward matter to
check that the expression ${\bf A} +{\bf bb}^T$ is the matrix
representation of the following twice-covariant non-singular
symmetric tensor at $\pi(n) \in {\mathcal B}$:
\begin{equation} \label{eq:symmetry13}
(\pi_*({\bf C}^{-1}))^{-1}.
\end{equation}
Note that the inversion inside the projection $\pi_*$ is on the
four-dimensional manifold $\mathcal M$, while the outer inversion is
on the three-dimensional manifold $\mathcal B$. Notice, too, that the
resulting three-dimensional covariant tensor is symmetric and
positive definite.

Having solved the preliminary scalar problem, we are now ready to
tackle the more delicate tensor identity (\ref{eq:symmetry7}).
Accordingly, we partition the matrix of components of the
body-time symmetric-tensor-valued function ${\tilde {\bf f}}$ as
follows:
\begin{equation} \label{eq:symmetry14}
{\tilde {\bf f}}({\bf C})= \left[
\begin{matrix}
{\bf M} & {\bf n} \\
{\bf n}^T & q
\end{matrix}
\right],
\end{equation}
where the dependence on the point $n \in \mathcal N$ has been
omitted for simplicity. Since we will need to evaluate this matrix
function also at ${\bf H}^T{\bf CH}$, we adopt the temporary
notation of indicating such quantities with a circumflex accent.
Thus, for instance, the $3 \times 3$ matrix  $\hat{\bf M}={\bf M}({\bf
H}^T{\bf CH})$ is the matrix taking the place of $\bf M$ in the
representation of ${\tilde {\bf f}}({\bf H}^T{\bf CH})$.
Performing all the matrix operations indicated in Equation
(\ref{eq:symmetry7}), we obtain the following identity:
\begin{equation} \label{eq:symmetry15}
\left[
\begin{matrix}
{\bf M} & {\bf n} \\
{\bf n}^T & q
\end{matrix}
\right] = \left[
\begin{matrix}
{\hat{\bf M}} & {\hat{\bf M}}{\bf h}+{\hat{\bf n}} \\
{\bf h}^T {\hat{\bf M}}+{\hat{\bf n}}^T & {\bf h}^T{\hat{\bf
M}}{\bf h}+{\bf h}^T{\hat{\bf n}}+{\hat {\bf n}}^T{\bf h}+{\hat q}
\end{matrix}
\right].
\end{equation}
We have at our disposal the degree of freedom of varying ${\bf h}$
to make this identity work. We start by observing that from the
upper-left block of the identity we obtain that:
\begin{equation} \label{eq:symmetry16}
{\bf M}({\bf C})={\bf M}({\bf H}^T{\bf CH}),
\end{equation}
where we have reverted to the standard notation for function
arguments. From our previous experience with the scalar law, we
conclude that the matrix $\bf M$ (that is, each of its entries)
will necessarily depend on $\bf C$ through the combination ${\bf
A} +{\bf bb}^T$. The remaining equations are:
\begin{equation} \label{eq:symmetry17}
{\bf n}({\bf C})={\bf M h}+{\hat {\bf n}},
\end{equation}
and
\begin{equation} \label{eq:symmetry18}
q({\bf C})={\bf h}^T{{\bf M}}{\bf h}+2{\bf h}^T {\hat{\bf n}}+
{\hat q},
\end{equation}
where the hatted quantities are to be evaluated at ${\bf H}^T{\bf
CH}$. The identity (\ref{eq:symmetry17}) is satisfied by the
following surprisingly simple form of $\bf n$:
\begin{equation} \label{eq:symmetry18a}
{\bf n}= {\bf Mb}+{\bf p},
\end{equation}
where $\bf p$ is an arbitrary function of ${\bf A}+{\bf bb}^T$.
Indeed, we have:
\begin{equation} \label{eq:symmetry18b}
{\hat {\bf n}}={\hat {\bf M}}{\hat{\bf b}}={\bf M} {\hat {\bf
b}}={\bf M}({\bf b}-{\bf h}),
\end{equation}
where we have made use of (\ref{eq:symmetry16}) and
(\ref{eq:symmetry10}). Plugging (\ref{eq:symmetry18a}) and
(\ref{eq:symmetry18b}) into (\ref{eq:symmetry17}), we convince
ourselves that the identity is indeed satisfied for all $\bf h$ by
the function ${\bf n}={\bf Mb}+{\bf p}$. We should also prove that
this is the {\it only} possible solution of this identity, but we
refrain from trying, considering ourselves fortunate to have found
this solution. As far as the remaining identity, Equation
(\ref{eq:symmetry18}), plugging into it the result
(\ref{eq:symmetry18a}) just obtained, and remembering that by
${\hat{\bf n}}$ we now mean ${\bf M}({\bf b}-{\bf h})$, we
conclude that it is satisfied by adopting the following form for
$q$:
\begin{equation}
q={\bf b}^T{\bf Mb}+2{\bf b}^T{\bf p}+r,
\end{equation}
where $r$ is an arbitrary function of ${\bf A}+{\bf bb}^T$. It can
be verified that, upon coordinate transformations in the body-time
bundle, the quantities $\bf M$, $\bf p$ and $r$ behave,
respectively, as a tensor, a vector and a scalar defined in the
base manifold. In a local frame consisting of $\bf U$ and three
vectors perpendicular to it (with respect to the pulled-back
metric $\bf C$), we have that ${\bf b}={\bf 0}$, and the blocks in
the matrix representing ${\tilde {\bf f}}$ are precisely $\bf M$,
$\bf p$, ${\bf p}^T$ and $r$.

Notice that the fact that we have restricted the argument $\bf C$
by means of the condition $C_{44}=-1$ (a condition that was used
repeatedly in the calculations) implies that the function ${\tilde
{\bf f}}({\bf C})$ can be determined only up to an additive term
of the form $\lambda {\bf U} \otimes {\bf U}$, as we already know.
This implies an indeterminacy in the choice of $r$.

\begin{rem} The condition (\ref{eq:symmetry18a}) is automatically
satisfied by any hyperelastic constitutive law such that $W=W({\bf
A}+{\bf b b}^T)$. Thus, every hyperelastic material whose (scalar)
constitutive law is reduced by the action of ${\mathcal H}^e$
gives rise to a stress-energy-momentum tensor also satisfying this
reduction. The converse, however, is not true in general: there
exist non-hyperelastic constitutive laws (for the
stress-energy-momentum tensor) that pass the test. Indeed, there
is no a-priori reason for $\bf M$ to be derivable from a
potential. In fact, if the constitutive law is hyperelastic then
it turns out that ${\bf p}={\bf 0}$ and that $\bf M$ is itself
derivable (according to the hyperelastic prescription
(\ref{eq:hyper2})) from a scalar function of ${\bf A}+{\bf bb}^T$.
\end{rem}

\begin{rem} We could have refrained from imposing ab initio the condition
$C_{44}=-1$, in which case we would have obtained an apparently
determinate component $r$. This determination, however, would have
only been illusory, since the general expression for the stress
contains a corresponding term affected by a Lagrange multiplier.
The situation is similar to what happens in classical Continuum
Mechanics when we deal with an internal constraint, such as
incompressibility. We may use an elastic constitutive law of a
compressible material which would provide a determinate
hydrostatic pressure (via, say, a bulk modulus) for any given
process, but this pressure is to be ultimately corrected by the
Lagrange multiplier. A different bulk modulus will certainly
change the pressure determined by the same process, but the
Lagrange multiplier will adapt its value to correct the situation.
\end{rem}

Our results so far can be summarized as follows. The
stress-energy-momentum tensor of a general relativistic elastic
material is given by the constitutive law:
\begin{equation} \label{eq:stress1}
{\bf t}= {\bf F} \;[\;{\tilde {\bf f}}(n, {\bf
C})\;+\;\lambda\;{\bf U} \otimes {\bf U}\;]\;{\bf F}^T,
\end{equation}
where ${\tilde {\bf f}}$ is a purely body-time tensor called the
Kirchhoff stress-energy-momentum. The Lagrange multiplier arises
from the requirement that all possible histories preserve the
metric structure of the fibres of the body-time manifold $\mathcal
N$. The strict dependence on the pulled-back metric $\bf C$ arises
from a requirement of relativistic frame indifference. Moreover,
although not mathematically necessary, one may require on physical
grounds that the symmetry group of elastic materials be large
enough to contain the group ${\mathcal H}^e$ consisting of all
local body-time transformations that preserve the body-projection
of vectors. In that case, the constitutive law is further reduced
so that (the determinate part of) the Kirchhoff
stress-energy-momentum in a local body-time frame ${\bf E}_\Gamma,
{\bf U}$ has the form:
\begin{eqnarray} \label{eq:stress2}
{\tilde {\bf f}}&=&M^{\Gamma \Delta} \;\;{\bf E}_\Gamma \otimes
{\bf E}_\Delta \nonumber \\ \; &+& \;(M^{\Gamma \Delta} \;\;
C_{\Delta 4}\;+\;p^\Gamma) \;\; ({\bf E}_\Gamma \otimes {\bf U} +
{\bf U} \otimes {\bf E}_\Gamma) \nonumber \\  \;&+&\;(M^{\Gamma
\Delta} \;\;C_{\Gamma 4}C_{\Delta 4}+2 C_\Gamma p^\Gamma + r)
\;\;{\bf U} \otimes {\bf U} \;,
\end{eqnarray}
where $M^{\Gamma \Delta}$, $p^\Gamma$ and $r$ are functions of the
twice-covariant body tensor:
\begin{equation} \label{eq:stress3}
(\pi_*({\bf C}^{-1}))^{-1}=(C_{\Gamma \Delta }+ C_{\Gamma 4} C_{4
\Delta})\;{\bf E}^\Gamma \otimes {\bf E}^\Delta.
\end{equation}
Finally, we will check that once this drastic reduction has taken
place and provided that ${\bf p}={\bf 0}$, the
stress-energy-momentum tensor has the 4-velocity vector $\bf u$ as
an eigenvalue. We check this fact directly by reverting to the
matrix notation. We start by ascertaining that $\bf U$ is an
eigenvector (with respect to the pulled-back metric $\bf C$) of
the total Kirchhoff stress-energy-momentum as follows:
\begin{equation} \label{eq:stress4}
\left[
\begin{matrix}
{\bf M} & {\bf Mb} \\
{\bf b}^T{\bf M} & {\bf b}^T{\bf Mb}+r+\lambda
\end{matrix}
\right] \left[
\begin{matrix}
{\bf A} & {\bf b} \\
{\bf b}^T & -1
\end{matrix}
\right] \left\{
\begin{matrix}
{\bf 0} \\
1
\end{matrix}
\right\}\;=\;\left\{
\begin{matrix}
{\bf 0} \\
 - (r+\lambda)
\end{matrix}
\right\},
\end{equation}
which proves the assertion. The fact that ${\bf u}={\bf FU}$ is an
eigenvector of $\bf t$ is now a straightforward consequence of the
definitions of $\bf t$ and $\bf C$. In fact, we could have
convinced ourselves directly of this result by simply adopting in
$\mathcal N$ a frame consisting of $\bf U$ and any three vectors
${\bf C}$-orthogonal to it, since under these conditions we have
${\bf b}={\bf 0}$ and, therefore ${\bf Mb}={\bf 0}$. From these
considerations it also follows that the (purely space-time)
stress-energy-momentum tensor has no mixed (time- with space-like)
components in a frame consisting of the 4-velocity vector $\bf u$
and any three space-like vectors $\bf g$-orthogonal to it.

Our purpose in this section has been to show that, except for the
presence of the Lagrange-multiplier term, all of the usual
assumptions about the form of the constitutive law in relativistic
elasticity, as pioneered by Carter and Quintana \cite{Car}, are
recovered in the present formulation by imposing a particular type
of body-time symmetry. Specifically, it must be assumed that the
symmetry group of the constitutive law of all materials under
consideration contains the group $\mathcal H^e$. It appears that,
although there is no strictly mathematical reason for adopting
such a restrictive criterion, there may be relatively strong
physical reasons to disregard all other materials. From the point
of view of the theory of anelasticity that we shall propose, the
question is not crucial. The main reason to have delved into the
elastic realm with such detail has been to make sure that the
four-dimensional body-time elastic archetype rests on a solid
foundation.

\section{Constitutive symmetries: solids and fluids}
\label{sec:symmetries}

We have already introduced and partially exploited the concept of
constitutive symmetry in Section \ref{sec:drastic}. Our purpose in
this section is to investigate the presence of further
constitutive symmetries, assuming that the constitutive law
already enjoys the standard symmetry introduced in Section
\ref{sec:drastic}. In other words, we assume that the symmetry
group ${\mathcal H}_n$ of the constitutive law at $n \in \mathcal
N$ contains ${\mathcal H}^e$ as a subgroup, as already expressed
in Equation (\ref{eq:symmetry6}). It is natural, therefore, to
introduce the quotient group
\begin{equation} \label{eq:materials1}
{\mathcal G}_n={\mathcal H}_n / {\mathcal H}^e
\end{equation}
as our object of interest and to call it the {\it reduced symmetry
group} of the constitutive law. Naturally, once a chart is chosen
in $\mathcal N$, this group is expressible as a multiplicative
matrix group. A general change of chart (always consistent with
the projection $\pi$ and with the structural group of the
body-time bundle) is given by four smooth functions $Y^1, Y^2,
Y^3, {\mathcal Y}^4$ of the form:
\begin{equation} \label{eq:materials2}
Y^\Lambda=Y^\Lambda(X^\Delta),\;\;\;\Lambda, \Delta = 1,2,3\;,
\end{equation}
and
\begin{equation} \label{eq:materials3}
Y^4=X^4+{\mathcal Y}^4(X^\Delta),\;\;\;\Delta=1,2,3.
\end{equation}
A moment's reflection reveals that, as far as the reduced symmetry
group is concerned, the function ${\mathcal Y}^4$ is irrelevant.
In fact, the study of the reduced symmetry group in terms of
coordinate representations boils down to the study of the
symmetries of the matrix function ${\bf M}({\bf A}+{\bf bb}^T)$.
Under coordinate transformations in the body-time principal
bundle, these symmetries are sensitive only to the part embodied
in Equation (\ref{eq:materials2}). Another way of expressing these
ideas is to say that we will investigate the symmetries of the
projected constitutive law on the base manifold $\mathcal B$. For
definiteness, but without much loss of generality, we shall
concentrate our attention on hyperelastic constitutive laws,
characterized, as we know, by a single scalar function W of a
purely material (three-dimensional) symmetric and positive
definite twice-covariant tensor ${\bf Z}={\bf A}+{\bf bb}^T$. A
(reduced) symmetry of such a constitutive law, consists of an
automorphism $\bf K$:
\begin{equation} \label{eq:materials4}
{\bf K}: T_{\pi(n)}{\mathcal B} \longrightarrow
T_{\pi(n)}{\mathcal B},
\end{equation}
such that:
\begin{equation} \label{eq:materials5}
W({\bf K}^T{\bf ZK})=W({\bf Z}),
\end{equation}
for all symmetric and positive definite twice-covariant tensors
$\bf Z$.

It is clear that the symmetry groups thus obtained for the same
constitutive law in two different charts are mutually conjugate,
the conjugation being established by the gradient of the change of
chart. In particular, if the symmetry group relative to one chart
is unimodular, then perforce it will be unimodular in all charts.
Bearing this idea in mind we may and shall adopt the standard
classification of material symmetries of classical Continuum
Mechanics. In particular, we have the following definitions:

\begin{defn} An elastic material point is a {\it solid} if the symmetry
group of its constitutive law as represented in any chart is
conjugate of a subgroup of the (Euclidean) orthogonal group.
\end{defn}

\begin{defn} An elastic material point is {\it isotropic} if its
symmetry group (in any chart) contains a conjugate of the
orthogonal group.
\end{defn}

\begin{defn} An elastic material point is a {\it fluid} if its
symmetry group is the unimodular group. \label{def:fluid}
\end{defn}

Notice that the unimodular group is characterized by the fact that
it is the largest subgroup of $GL(3,\mathbb{R})$ that preserves
all volume forms. In any given chart, this implies that the
function $W$ depends on its argument (namely, ${\bf Z}$) through
the determinant of $\bf Z$. In order for this function to be
invariant under changes of coordinates, therefore, we need to
introduce any other volume form (for instance, the baryon form)
and make $W$ depend on the ratio between the determinant of $\bf
Z$ and the component of this form. We conclude that the most
general constitutive equation of a fluid is given by an arbitrary
function of this ratio.

Let the quotient group ${\mathcal G}_n$ be known. Then the
corresponding original group ${\mathcal H}_n$ consists of the $4
\times 4$-matrices of the form:
\begin{equation} \label{eq:materials6}
{\bf H} = \left[
\begin{matrix}
{\bf K} & {\bf 0}  \\
{\bf h}^T & 1
\end{matrix}
\right],
\end{equation}
where ${\bf K}$ is in ${\mathcal G}_n$ and ${\bf h}$ is arbitrary.
In other words, if a constitutive law expressed in terms of $\bf
Z$ satisfies (\ref{eq:materials5}), then the corresponding
constitutive law in terms of $\bf C$ as an independent variable
will satisfy a similar equation, namely:
\begin{equation} \label{eq:materials7}
W({\bf H}^T{\bf CH})=W({\bf C}),
\end{equation}
where $\bf H$ is an arbitrary matrix of the form
(\ref{eq:materials6}). The verification of this property and of
the fact that these matrices transform in the appropriate way
under coordinate transformations of the type (\ref{eq:materials2})
and (\ref{eq:materials3}) is a straightforward exercise.

\section{Constitutive isomorphisms and uniformity}

A material body-time complex ${\mathcal N}$ may enjoy symmetries
that go beyond the constitutive symmetries of each of its points.
These ``symmetries'', arising from the comparison of the
constitutive responses at different points of the body-time
manifold, confer to it an extra geometrical structure, namely,
that of a Lie groupoid. In classical Continuum Mechanics it is the
time evolution of this entity that allows for a rigorous
theoretical formulation of the anelastic behaviour characteristic,
for example, of ideal elasto-plasticity and of bulk growth. In
this section we will review some of the basic notions of this
theory as they apply within the context of General Relativity. To
emphasize the fact that these notions are quite independent of
other concepts introduced so far (such as the various reductions
of the constitutive law), we will return to the primeval form of a
constitutive law for some space-time (scalar-, vector- or
tensor-valued) constitutive function $\Psi$, namely:
\begin{equation} \label{eq:uniformity1}
\Psi =\Psi(n, {\bf g}, {\bf F}).
\end{equation}
\begin{defn} Two points $n_1, n_2 \in \mathcal N$ are {\it
constitutively isomorphic} if there exists an isomorphism ${\bf
P}_{12}$ between their tangent spaces:
\begin{equation} \label{eq:uniformity2}
{\bf P}_{12}: T_{n_1}{\mathcal N} \longrightarrow T_{n_2}{\mathcal
N},
\end{equation}
such that the equation
\begin{equation} \label{eq:uniformity3}
\Psi(n_2, {\bf g}, {\bf F})=\Psi(n_1, {\bf g}, {\bf FP}_{12})
\end{equation}
is satisfied identically for all $\bf g$ and for all $\bf F$ in
their respective domains of existence.
Such a ${\bf P}_{12}$ is said to be a \emph{constitutive isomorphism}.
For consistency, we will assume that the isomorphism ${\bf P}_{12}$ respects the unit
vector field ${\bf U}$, namely:
\begin{equation} \label{eq:uniformity3a}
{\bf P}_{12} {\bf U}(n_1)= {\bf U}(n_2).
\end{equation}
In local charts consistent with a trivialization of the body-time
complex $\mathcal N$, this condition is equivalent to ${\bf
P}_{12}$ having a matrix representation of the form
(\ref{eq:symmetry3}).
\end{defn}

The preceding definition is the relativistic analog of the
classical notion of material isomorphism introduced by Noll
(\cite{NLFT, Noll}) It is clear that this definition reverts to
that of a constitutive symmetry, namely (\ref{eq:symmetry}), if we
just identify $n_1$ with $n_2$. From the physical point of view,
we may say that when two points are constitutively isomorphic they
are ``made of the same material''. Indeed, when two points are
constitutively isomorphic, there exist local charts that render
the coordinate expressions of their constitutive laws identical to
each other, and vice-versa.

\begin{defn} A material body-time complex $\mathcal N$ is said to
be {\it constitutively uniform} (or, simply, {\it uniform}) if its
points are pair-wise constitutively isomorphic. If the material
isomorphisms can be chosen to depend smoothly on both the source
and the target points, $\mathcal N$ is said to be {\it smoothly
uniform}.
\end{defn}

Due to the assumed fibred nature of the body-time manifold, it is
also possible to introduce the following notion.

\begin{defn} A body-time manifold $\mathcal N$ is {\it time-wise
uniform} if the points within each fibre are pair-wise
constitutively isomorphic.
\end{defn}

Physically, a time-wise uniform body-time manifold is such that
each point in the base manifold $\mathcal B$ preserves its own
constitutive nature (its ``chemical identity'', as it were) as
time goes on, without regard as to whether this nature is
comparable with that of any other point of $\mathcal B$. Clearly,
\begin{equation} \label{eq:timewise}
{\rm uniformity} \Rightarrow {\rm time-wise \; uniformity},
\end{equation}
but not conversely. Because of the ``one-sided'' bundle nature of
the material body-time, there is no spatial counterpart to the
concept of time-wise uniformity. On the other hand, if a body-time complex
is known to be time-wise uniform and if for each point of the base
manifold there exists a local cross-section the points of whose
image are constitutively isomorphic, then the body-time complex is
necessarily uniform.

Suppose that $\mathcal N$ is smoothly uniform. It is clear that if
${\bf P}_{12}$ is a constitutive isomorphism between the source
point $n_1$ and the target point $n_2$, then the inverse ${\bf
P}_{21} = {\bf P}^{-1}_{12}$ is a constitutive isomorphism in
which the source and the target have been exchanged. Moreover, if
${\bf P}_{12}$ and ${\bf P}_{23}$ are constitutive isomorphisms
between $n_1$ and $n_2$ and between $n_2$ and $n_3$, respectively,
then the composition ${\bf P}_{13}={\bf P}_{23} {\bf P}_{12}$ is a
constitutive isomorphism between $n_1$ and $n_3$. Finally, the
collection of all constitutive automorphisms at each point $n$
constitutes a group (more precisely, this is the constitutive
symmetry group ${\mathcal H}_n$ at that point). It is not
difficult to prove that the symmetry groups at two different
points are conjugate of each other, and that the conjugation is
achieved by any constitutive isomorphism between these two points.
Conversely, given a constitutive isomorphism ${\bf P}_{12}$
between $n_1$ and $n_2$, the totality ${\mathcal P}_{12}$ of
constitutive isomorphisms between these points is given by:
\begin{equation} \label{eq:uniformity4}
{\mathcal P}_{12}={\bf P}_{12} {\mathcal H}_{n_1}={\mathcal
H}_{n_2} {\bf P}_{12}.
\end{equation}
All these properties taken together confer upon a uniform
body-time complex $\mathcal N$ the structure of a transitive Lie
groupoid.

A local chart $(X^A)$ on $\mathcal N$ whose domain contains $n$ and $n^\prime$
\emph{naturally} induces the map ${\bf O} : T{\mathcal N} \longrightarrow
T{\mathcal N}$ which written as a two-point tensor has the form
\begin{equation}
{\bf O}(n,n^\prime) = \frac{\PD}{\PD X^A}\biggm|_n\otimes\,\, dX^A\biggm|_{n^\prime}.
\end{equation}
This gives rise to: 
\begin{defn} A smoothly uniform body-time complex $\mathcal N$ is
{\it locally homogeneous} if, for each point $n \in \mathcal N$,
there exists a chart (containing $n$) such that the maps it
naturally
induces between tangent spaces are constitutive isomorphisms.
\end{defn}

It is often useful to exploit the transitive character of the
material isomorphisms to extract a particular point,
$n_0$ say, from $\mathcal N$ and to use it as an {\it archetype}
in the sense that the definition of uniformity is equivalent to
the following fact: all points of $\mathcal N$ are constitutively
isomorphic to this archetype. Denoting a material isomorphism from
the archetype $n_0$ to a point $n$ by ${\bf P}(n)$, and the
corresponding set of all such isomorphisms by ${\mathcal P}(n)$,
we see that choosing a frame $\{{\bf E}^{(0)}_A\}$ at the archetype
induces at each point $n$ the collection of frames $\{{\bf E}_A(n) = {\mathcal
P}(n){\bf E}^{(0)}_A\}$, which is a subset of all the possible frames at
that point. This subset is governed by the constitutive symmetry
of the archetype.

\newcommand{\BodyDel}{\hat{\nabla}}
In the case where the symmetry group is the identity each of
the sets ${\mathcal P}(n)$ contains only ${\bf P}(n)$.
For each such ${\bf P}(n)$ and frame $\{{\bf E}_A(n)\}$
one may define a linear connection $\BodyDel$ on $\mathcal N$
such that $(\BodyDel {\bf E}_A)(n) = 0$.
In this frame the components $\{P^A_B\}$ of
\begin{equation}
{\bf P}(n) = P^A_B(n,n_0){\bf E}_A(n)\otimes{\bf B}^B(n_0)
\end{equation}
are constants, where $\{{\bf B}^A\}$ is the dual coframe field
to $\{{\bf E}_A\}$
i.e. ${\bf B}^A({\bf E}_B) = \delta^A_B$ for all $n$.
The existence of the $\BodyDel$-parallel frame field $\{{\bf
E}_A\}$ implies that the curvature of $\BodyDel$ vanishes. Furthermore,
if the $\{{\bf E}_A\}$ are holonomic, i.e. ${\bf E}_A=\frac{\PD}{\PD X^A}$ in
some local chart $X^A$, then the torsion of $\BodyDel$ also vanishes
and $\BodyDel$ is said to be \emph{flat}. Should the torsion of
$\BodyDel$ not vanish, one can say that $\mathcal N$ contains a smooth
distribution of inhomogeneities or dislocations. More generally, the
sets $\mathcal P$ contain more than one member and their geometric
interpretation is formulated in terms of $G$-structures \cite{EES1,EL1,EL2}. 
The notion of local homogeneity corresponds exactly to a notion of
flatness of these $G$-structures.

A smoothly uniform body-time complex $\mathcal N$ (whether locally
homogeneous or not) can be described in terms of the archetypal
constitutive law ${\bar {\Psi}}({\bf g}, {\bf F})=\Psi(n_0, {\bf
g}, {\bf F})$ of a point $n_0$ in the form:
\begin{equation} \label{eq:uniformity5}
\Psi(n, {\bf g}, {\bf F})={\bar \Psi}( {\bf g}, {\bf FP}(n)).
\end{equation}
Assume now that the archetypal constitutive law happens to be
hyperelastic and that we identify $\Psi$ with the elastic energy
$W$. Regarding, accordingly, $W$ as a function of $\bf g$, $\bf F$
and ${\bf P}(n)$, we obtain, by virtue of Equation
(\ref{eq:uniformity5}):
\begin{equation} \label{eq:uniformity6}
\frac{\partial W}{\partial {\bf P}}={\bf F}^T\frac{\partial
W}{\partial {\bf F}}{\bf P}^{-T},
\end{equation}
or, invoking (\ref{eq:hyperelastic1}):
\begin{equation} \label{eq:uniformity7}
\frac{\partial W}{\partial {\bf P}}=[-W {\bf I}+{\bf F}^T{\bf
f}\;{\bf F}^{-T}] \;{\bf P}^{-T}.
\end{equation}
which, in components, reads:
\begin{equation} \label{eq:uniformity8}
\frac{\partial W}{\partial P^A_M}=[-W \delta^B_A+g_{ad}\;F^d_A\;
f^{ab}\;(F^{-1})^B_b] \;(P^{-1})^M_B.
\end{equation}

The quantity enclosed in square brackets is a purely body-time
tensor which, following the classical counterpart \cite{Ep0, EM1,
EM2}, is referred to as the determinate part of the {\it
relativistic Eshelby tensor} . From the above formula, it is clear
that its physical meaning is related to the elastic energy
expended in producing a change (or remodeling) of a first-order
neighbourhood in the body-time manifold. In view of other concepts
introduced in this section, this interpretation can be worded as
follows: the amount of elastic energy required to change the
pattern of distribution of inhomogeneities.

If the constitutive equation is given explicitly in terms of an
equation such as (\ref{eq:uniformity5}), with a specific choice of
the field ${\bf P}(n)$ (with the degree of freedom afforded by the
symmetry group, of course), it means that we have somehow been
able to specify the constitutive behaviour at each point of the
base manifold for all times. In practice, however, starting from
an initial space-like Cauchy manifold, the constitutive law will
evolve at each point in a way that, although always abiding by
Equation (\ref{eq:uniformity5}), is determined by some extra
constitutive criterion that allows a specific ${\bf P}$ to be
pinned down according to, for example, the local value of the
Eshelby tensor. Thus, the $\bf P$-maps function somewhat as
internal state variables governed by some laws of evolution.

\section{Evolution laws}
\label{sec:evol}

Let $\mathcal R$ denote the canonical right
action of $\mathbb{R}$ on $\mathcal N$. For each $s \in
\mathbb{R}$ we have, therefore, a fibre-preserving diffeomorphism:
\begin{eqnarray} \label{eq:evol1}
\mathcal{R}_s\;:\; {\mathcal N} &\longrightarrow& {\mathcal N}
\nonumber \\
n &\mapsto& \mathcal{R}_s(n) \equiv ns
\nonumber\\
(X^\Lambda,S) &\mapsto& (X^\Lambda,S+s)
\end{eqnarray}
The induced tangent map:
\begin{equation} \label{eq:evol2}
\mathcal{R}_{s_*}\;:\;T_s\mathcal{N} \longrightarrow
T_{ns}\mathcal{N}
\end{equation}
is, therefore, a non-singular linear map between the corresponding
tangent spaces. Conversely, given two points, $n_1,n_2 \in
\mathcal{N}$, such that $\pi(n_1)=\pi(n_2)$, there exists a unique
$s \in \mathbb{R}$ such that $n_2=n_1s$ and, therefore, a uniquely
determined map ${\bf R}_{12}$ between their tangent spaces. Assume
now that we are given a time-wise uniform body complex
$\mathcal{N}$ and that for each pair of points $n_1$ and $n_2$
that lie in a given fibre $\pi^{-1}(b)$ the map ${\bf R}_{12}$
just defined happens to be a constitutive isomorphism. In such a
case we are justified in saying that, as far as the point $b \in
\mathcal{B}$ is concerned, the constitutive equation {\it does not
evolve in time}. If this is the case for all points of $\mathcal
{B}$, we say that the constitutive equation of $\mathcal{N}$ is
{\it non-evolutive}. Conversely, if for at least one point $b$ of
$\mathcal{B}$ and at least one pair of points $n_1, n_2 \in
\pi^{-1}(b)$ the right-action-induced map ${\bf R}_{12}$ is not a
constitutive isomorphism, we are in the presence of {\it material
evolution} \cite{Ep1,Ep2}.

As pointed out in Section \ref{sec:body-time}, we can associate
with the one-parameter group action $\mathcal{R}$ the fundamental
unit vector field $\bf U$. Let us denote the Lie derivative with
respect to $\bf U$ by means of a superposed dot, namely, for a
vector field $\bf V$:
\begin{equation} \label{eq:evol3}
\dot{\bf V}\equiv L_{\bf U}{\bf V} = [{\bf U}, {\bf V}].
\end{equation}
Let $\mathcal{N}$ be a time-wise uniform body-time complex. On
each fibre $\pi^{-1}(b)$ we may, therefore, choose a particular
point, $n_0(b)$ say, as a {\it fibre archetype} and denote by
${\bf P}(n)$ a smooth choice of constitutive isomorphisms from
$n_0(b)$ to the points $n$ lying on the corresponding fibre. These
linear maps, being two-point tensors, can be regarded as vector
fields (more precisely, as vector-field-valued covectors at the
archetype). If the body-time is actually uniform, then a single
archetype can be chosen. The non-evolution condition can then be
stated as:
\begin{equation} \label{eq:evol4}
\dot{\bf P}={\bf 0}.
\end{equation}
Accordingly, an {\it evolution law} at a point $n \in \mathcal{N}$
will be an expression of the form:
\begin{equation} \label{eq:evol5}
\dot{\bf P}={\bf \Phi}({\bf P}, {\bf e}, ...; n),
\end{equation}
where $\bf e$ is the (determinate part of the) relativistic
Eshelby tensor (as per Equations (\ref{eq:uniformity7}) or
(\ref{eq:uniformity8})), and where other arguments could be
included. As is the case in the non-relativistic counterpart, laws
of evolution are subjected to a number of formal restrictions,
some of which we will presently derive.

The first restriction on the possible forms of the evolution
function $\bf \Phi$ stems from the uniformity requirement itself.
Indeed, if all the points are made of the same material, the
evolution law of each point $c$ should be obtained as the push
forward by ${\bf P}(n)$ of the evolution law at the archetype,
namely, a law of the form $\dot{\bf P} = {\bf \Phi}_0( {\bf e}_0
)$. As a result of this requirement, we obtain that the evolution
law must necessarily be of the form:
\begin{equation} \label{eq:evol6}
\dot{\bf P} ={\bf P}\; {\bf \Phi}_0({\bf P}^T {\bf e} {\bf
P}^{-T}).
\end{equation}
Introducing the notation:
\begin{equation} \label{eq:evol7}
{\bf L}_{P_0}={\bf P}^{-1}\dot{\bf P},
\end{equation}
we write the reduced evolution law as:
\begin{equation} \label{eq:evol8}
{\bf L}_{P_0} = {\bf \Phi}_0({\bf P}^T {\bf e} {\bf P}^{-T}).
\end{equation}
The quantity:
\begin{equation} \label{eq:evol9}
{\bf L}_P=\dot{\bf P} {\bf P}^{-1},
\end{equation}
can be referred to, by abuse of terminology, as the {\it (general
relativistic) inhomogeneity velocity gradient}. It is a linear map
of the tangent space $T_n{\mathcal N}$ into itself. The tensor
${\bf L}_{P_0}={\bf P}^{-1} {\bf L}_P {\bf P}$ is, therefore, the
pull-back of the inhomogeneity velocity gradient to the archetype.
To summarize, the reduction of the general evolution law by
uniformity arguments leads to an explicit dependence on the
uniformity map ${\bf P}(n)$, as given in Equation
(\ref{eq:evol6}). It is worthwhile noting that in a local chart
consistent with a trivialization of the body-time complex, the
matrix representing the tensor ${\bf L}_P$ has the generic form:
\begin{equation} \label{eq:evol19a}
{\bf L}_P = \left[
\begin{matrix}
{\bf L} & {\bf 0}  \\
{\bf m}^T & 0
\end{matrix}
\right],
\end{equation}
where ${\bf L}$ and ${\bf m}$ are, respectively, an arbitrary
square matrix and an arbitrary vector of order 3.

Since the uniformity maps are in general non-unique, and since
their lack of uniqueness is governed by the constitutive symmetry
group, it is to be expected that further reductions (due to
symmetry) will be possible whenever the symmetry group is non
trivial. These reductions, to whose derivation we turn presently,
are of two kinds.

The first symmetry reduction, called the {\it principle of actual
evolution} \cite{EM3, Ep2}, stems from the observation that the
condition (\ref{eq:evol4}) is in fact sufficient but not necessary
for claiming that the constitutive law does not evolve. Indeed,
let ${\bf P}(n)$ be a smooth choice of constitutive isomorphisms
from $n_0(b)$ to the points $n$ lying on the fibre $\pi^{-1}(b)$,
and let this ${\bf P}(n)$ satisfy (\ref{eq:evol4}). Consider now a
different choice of constitutive isomorphisms given by:
\begin{equation} \label{eq:evol10}
{{\bf Q}}(n)={\bf P}(n){\bf G}(n),
\end{equation}
where ${\bf G}(n)$ is a smooth one-parameter ($n$) family of
material symmetries of the fibre archetype such that ${\bf
G}(n_0)={\bf I}$. Such a choice can be made non-trivially provided
that the symmetry group is continuous. Clearly, the constitutive
isomorphisms ${\bf Q}(n)$ represent the same material phenomenon
as the original ${\bf P}(n)$, since they differ at any point along
the fibre just by a material symmetry of the archetype. In other
words, the choice ${\bf Q}(n)$ corresponds to a non-evolving
situation along the fibre. Nevertheless, ${\bf Q}(n)$ will in
general fail to satisfy condition (\ref{eq:evol4}). Indeed, taking
the Lie-derivative of Equation (\ref{eq:evol10}) with respect to
$\bf U$ we obtain:
\begin{equation} \label{eq:evol11}
{\bf Q}^{-1} \dot{\bf Q} ={\bf G}^{-1} \dot{\bf G},
\end{equation}
where we have made use of (\ref{eq:evol4})\footnote{Note that
$\dot{\bf G}$ coincides with the ordinary derivative with respect
to the (real) fibre parameter.}. It follows then that, as long as
the evolution function ${\bf \Phi}_0$ gives a result within the
Lie algebra of the archetype, there is no evolution. The principle
of actual evolution, therefore, states that the evolution law must
have values lying outside of the Lie algebra of the archetype. Two
apparently different evolution laws whose results differ by an
element of this Lie algebra are, therefore, to be considered as
equivalent.

Finally, the evolution law must be invariant under the action of
the symmetry group of the archetype, namely:
\begin{equation} \label{eq:evol12}
{\bf \Phi}_0({\bf e}_0 )={\bf G}\; {\bf \Phi}_0({\bf G}^T {\bf
e}_0 {\bf G}^{-T}) \; {\bf G}^{-1} + \mathbb{G},
\end{equation}
for all members $\bf G$ of the symmetry group. In this formula,
$\mathbb{G}$ is an arbitrary element of the Lie algebra of the
symmetry group of the archetype. The reason for the presence of
this term should be clear from the principle of actual evolution.

\section{Example: an anelastic relativistic fluid}

According to Definition \ref{def:fluid} and the remarks
thereafter, a hyperelastic fluid point is completely characterized
by a scalar function of the form:
\begin{equation} \label{eq:example1}
W=W(\sqrt{\det({\bf Z})}/\omega),
\end{equation}
\newcommand{\Intder}[1]{\iota_{#1}}
where $\omega$ is the component of some given 3-volume form $\# 1$ on
the three-dimensional base manifold $\mathcal{B}$:
\begin{equation}
\# 1 = \omega\, dX^1 \wedge dX^2 \wedge dX^3
\end{equation}
in the chart $X^\Gamma$.
We recall that, in accordance with the notation of Section
\ref{sec:symmetries}, $\bf{Z}=\bf{A}+\bf{bb}^T$ is the matrix
representation of the tensor $(\pi_*({\bf C}^{-1}))^{-1}$, defined
at each point of the three-dimensional base manifold
$\mathcal{B}$. The above
expression $\sqrt{\det({\bf Z})}/\omega$, though,
is chart independent because it is the ratio between the
components of two tensor densities of the same type.
It is now a straightforward matter to obtain the
most general expression for the stress-energy-momentum tensor
${\bf t}$ corresponding to a potential $W$ of the form
(\ref{eq:example1}) through the use of Equations
(\ref{eq:reduced}) and (\ref{eq:hyper2}). The calculation is
greatly facilitated by the fact that:
\begin{equation} \label{eq:example2}
\det({\bf Z})=-\det({\bf C}),
\end{equation}
as it follows from Equation (\ref{eq:symmetry9})\footnote{Perhaps
the easiest way to convince oneself of this fact is to multiply
the matrix (\ref{eq:symmetry9}) to the left by the matrix
$\label{eq:footnote} \left[
\begin{matrix}
{\bf I} & {\bf b} \\
0 & 1
\end{matrix}
\right] ,
$
whose determinant is clearly $1$.}. In other words, for the case
of a fluid, it doesn't matter whether the space-time volume form
is defined from the fully-fledged pull-back of the spacetime
metric or from its projection on the base manifold. This
conclusion is also consistent with the fact that the determinant
of $\bf{C}$ automatically passes the drastic-reduction test
embodied in Equation (\ref{eq:symmetry8}).
Such considerations indicate that the ratio $\sqrt{\det({\bf
Z})}/\omega$ can be written in a coordinate-free manner. In fact
\begin{equation}
\sqrt{\det({\bf Z})}/\omega = \#^{-1}\left[\Intder{{\bf U}}\left(\kappa^*\star 1\right)\right]
\end{equation}
where $\Intder{{\bf U}}$ is the interior derivative on forms, 
$\star 1$ is the volume form of the spacetime manifold $\mathcal{M}$
and $\#^{-1}$ is the inverse of the the Hodge map $\#$ associated with $\# 1$. 
Recalling the formula
for the derivative of a determinant, namely,
\begin{equation}
\frac{\partial \det ({\bf C})}{\partial {\bf C}}=\det({\bf C})
{\bf C}^{-T},
\end{equation}
the final form of the stress-energy-momentum tensor is obtained
as:
\begin{equation} \label{eq:example3}
{\bf t} = p(\sqrt{\det(\bf{Z})}/\omega) \; {\bf g}+ \lambda {\bf
u} \otimes {\bf u}
\end{equation}
as expected, where $p$ is a scalar function of its argument. More
precisely, using Equation (\ref{eq:hyperelastic1}), we obtain that
$p$ is related to $W$ via the formula:
\begin{equation} \label{eq:example3a}
p=W+\;\frac{\sqrt{\det ({\bf Z})}}{\omega}\;W',
\end{equation}
where $W'$ indicates the derivative of the function $W$ of
Equation (\ref{eq:example1}) with respect to its argument.

Assume now that we are given a uniform body-time complex $\mathcal
N$ modelled after a given constitutive law of the type
(\ref{eq:example3}) for the archetype. This means that for each
point $n \in \mathcal N$ there exists a linear map ${\bf P}(n)$
from the tangent space $T_{n_0}\mathcal N$ at the archetype $n_0$ to
$T_n\mathcal N$ such that the
stress-energy-momentum tensor at $n$ is given by:
\begin{equation} \label{eq:example4}
\begin{split}
{\bf t} &= p\left[\sqrt{\det({\bf P}{\bf Z}{\bf P}^T)}/\omega
\right] \; {\bf g}+ \lambda {\bf u} \otimes {\bf u},\\
&= p\left[(\det({\bf P}))\sqrt{\det({\bf Z})}/\omega
\right] \; {\bf g}+ \lambda {\bf u} \otimes {\bf u}
\end{split}
\end{equation}
where $\bf Z$ is now evaluated through the pull-back by $\kappa$
of $\bf g$ to $n$ and $\det({\bf P})$ is the component of the two-point tensor
\begin{equation}
\det({\bf P}) \left(\frac{\PD}{\PD X^1} \wedge \frac{\PD}{\PD X^2} \wedge
\frac{\PD}{\PD X^3} \wedge
\frac{\PD}{\PD S}\right)\biggm|_n \otimes \left(dX^1\wedge
dX^2\wedge dX^3\wedge dS\right)\biggm|_{n_0} 
\end{equation}
induced by ${\bf P}$.
Equivalently, we can write the constitutive
equation (\ref{eq:example4}) in terms of ${\bf C}$ at $n$ as:
\begin{equation} \label{eq:example5}
\begin{split}
{\bf t} &= p\left[(\det({\bf P}))\sqrt{-\det({\bf C})}/\omega
\right] \; {\bf g}+ \lambda {\bf u} \otimes {\bf u}.
\end{split}
\end{equation}
 
In fact, we can look at this constitutive equation as specifying
at each point $n \in \mathcal N$ a behaviour of the type
(\ref{eq:example3}), but with a reference volume two-point form defined
locally by the formula:
\begin{equation} \label{eq:example6}
\frac{\omega}{(\det({\bf P}))}\; dX^1\wedge dX^2 \wedge dX^3
\wedge dS.
\end{equation}
Notice that in this formula $\omega$ is the component of $\# 1$ at the
archetype $n_0$, so that (\ref{eq:example6}) is
the contraction of the $4$-form $\# 1\wedge dS$ (at the archetype $n_0$) with the two-point tensor
\begin{equation}
\det({\bf P}^{-1}) \left(\frac{\PD}{\PD X^1} \wedge \frac{\PD}{\PD X^2} \wedge
\frac{\PD}{\PD X^3} \wedge
\frac{\PD}{\PD S}\right)\biggm|_{n_0} \otimes \left(dX^1\wedge
dX^2\wedge dX^3\wedge dS\right)\biggm|_n
\end{equation}
induced by ${\bf P}^{-1}$.

So as to construct a model of an anelastic fluid fashioned after
the archetypal constitutive law (\ref{eq:example5}), we will allow
the referential volume form (\ref{eq:example6}) to evolve
according to an evolution law of the type (\ref{eq:evol8}). For
the case of a fluid, the principle of actual evolution (introduced
in Section \ref{sec:evol}), stipulates that the function $\Phi_0$
in Equation (\ref{eq:evol8}) must not be of the form:
\begin{equation} \label{eq:example7}
\left[
\begin{matrix}
{\bf K}' & {\bf 0}  \\
{\bf h'}^T & 0
\end{matrix}
\right].
\end{equation}
where ${\bf h'}$ is arbitrary and ${\bf K}'$ is  traceless
(because the Lie algebra of the unimodular group is precisely the
algebra of traceless matrices).

As far as the restrictions placed by Equation(\ref{eq:evol12}), we
start by noting that the determinate part of the relativistic
Eshelby tensor is given, according to Equations
(\ref{eq:uniformity8}), (\ref{eq:example1}) and
(\ref{eq:example3}), by:
\begin{equation} \label{eq: example8}
{\bf e}=(p-W)\;{\bf I}.
\end{equation}
The restrictions just mentioned are satisfied by the following
evolution equation expressed in matrix form as:
\begin{equation} \label{eq:example9}
{\bf P}^{-1}{\dot {\bf P}}= \left[
\begin{matrix}
\phi(p-W){\bf I} & {\bf 0}  \\
{\bf 0}^T & 0
\end{matrix}
\right],
\end{equation}
where $\phi$ is an arbitrary scalar-valued function of a scalar
argument. Naturally, this formula is not form-invariant
under (trivialization-consistent) coordinate transformations.
Nevertheless, the extra terms that would appear affect only the
first three entries of the fourth row, which is permitted
according to the principle of actual evolution. In other words,
the offending terms would belong to the Lie algebra of the group
${\mathcal H}^e$.

At this point, it may prove worthwhile to indicate the
non-relativistic counterpart of the example we have proposed.
Consider a compressible elastic fluid which cannot sustain any
shear stress, so that its constitutive law consists simply of a
hydrostatic stress $p$ determined by the ratio between its present
volume $V$ and a given reference volume $V_0$. A possible
constitutive law is: $p=k(V/V_0 -1)$, where $k$ is positive
material constant measuring the elasticity of the material. If,
for example, this material were to be squeezed into a rigid
container of a volume smaller than $V_0$, it would sustain a
compression that would persist as long as the material and the
container are left unaltered. If, however, the reference volume
(or resting volume) were to begin to decrease with time, the
pressure would accordingly decrease. This is a volumetric version
of the well-known phenomenon of stress relaxation, observed to
varying degrees in all real materials subjected to a fixed
uniaxial extension. The rate of decrease of the reference volume
may be directly related to the pressure, so that as the reference
volume approaches the volume of the container, the pressure
approaches zero and the process tends to fade asymptotically. A
possible evolution law reads: ${\dot V}_0=hp$, where $h$ is a
positive material constant representing the relaxation properties
of the material. In the relativistic picture that we have
presented, the role of $V$ is played by $\sqrt{- \det {\bf C}}$,
while $V_0$ is represented by $\omega/ (\det {\bf P})$.

\bigskip

{\bf Acknowledgment} This work has been partially supported by the
Natural Sciences and Engineering Research Council of Canada.

\section{Appendix}

An alternative approach to determine the structure of ${\bf t}$
is via a variational principle. The previous discussion is formulated
in terms of the embedding map $\kappa$ from the body-time manifold
$\mcN$ to the spacetime manifold $(\mcM,\Bg)$. To construct a
variational principle it proves expedient to employ the inverse
$f\equiv\kappa^{-1}$ of $\kappa$,
\begin{equation}
\begin{split}
f : \mcM &\rightarrow \mcN\\
m &\mapsto n=\kappa^{-1}(m).
\end{split}
\end{equation}

Let $(\mcE,\pi_{\mcE\mcM},\mcM)$ be a bundle with fibre $\mcN$, and
$(x^a,X^A)$ be the coordinates of a point in $\mcE$ with the base
point $(x^a)$, i.e. $x^a=\pi^a_{\mcE\mcM}(x,X)$. Thus, $f$
specifies a section $\sigma$ of $\mcE$ with the coordinate representation
\begin{equation}
\begin{split}
\sigma : \mcM &\rightarrow \mcE\\
x^a &\mapsto (x^a,f^A(x)).
\end{split}
\end{equation}
Let $J^1\mcE$ be the first jet bundle of $(\mcE,\pi_{\mcE\mcM},\mcM)$
with fibre $F_p$ at $p\in\mcM$ and $(X^A,\mcF^A_a)$ coordinate
points in $F_p$. Take $(\gamma^{ab}$) to be coordinates on
$\Omega_p\subset T_p\mcM\otimes T_p\mcM$, 
the space of non-degenerate rank 2 symmetric contravariant tensors at
$p\in\mcM$.
Let $\mathbb R_p$ be the real line at $p\in\mcM$ and coordinate it by $(l)$.
The variational principle below is expressed in terms
of local sections of a fibre bundle $(\mcD,\pi_{\mcD\mcM},\mcM)$ whose fibre
over $p\in\mcM$ is $F_p\times\Omega_p\times\mathbb R_p$. Thus, a point
in $\mcD$ is coordinated by $(x^a,X^A,\mcF^A_a,\gamma^{ab},l)$.

The inverse metric tensor field $\BG\equiv {\bf g}^{-1}$
on $\mcM$ and the scalar field $\lambda$ on
$\mcM$ are used to ``prolong'' $\sigma$ to the section $\SecD$ of
$\mcD$: 
\begin{equation}
\begin{split}
\SecD : \mcM &\rightarrow \mcD\\
x^a &\mapsto (x^a,f^A(x),\PD_a f^A(x),G^{ab}(x),\lambda(x))
\end{split}
\end{equation}
where $G^{ab}\equiv \BG(dx^a,dx^b)=\Bg^{-1}(dx^a,dx^b)$
are the components of $\Bg^{-1}$ with respect $(x^a)$ on $\mcM$.

The action functional has the form
\begin{equation}
\label{appendix:action_functional}
\mcS[\Gamma] = \int_{\Gamma[\mcM]}\rho\Bomega
\end{equation}
where $\Gamma=\SecD$ and $\Bomega$ is the naturally induced volume $4$-form
\begin{equation}
\label{appendix:omega_form}
\Bomega = \frac{1}{\sqrt{-\det[\gamma^{ab}]}}
dx^1\wedge dx^2\wedge dx^3\wedge dx^4
\end{equation}
on $\mcD$ and $\rho$ is a $0$-form on $\mcD$. 

We consider here media where the $0$-form $\rho$ is constructed out of functions on
$\mcD$ including the components of a $1$-form $\Bmu=\mu_A(X) dX^A$ on $\mcN\subset\mcD$:
\begin{equation}
\label{appendix:rho_in_terms_of_W}
\rho = W(X,\mcF,\gamma) -
\frac{1}{2}l\left(\mu_A(X)\mu_B(X)\mcF^A_a\mcF^B_b\gamma^{ab} +
1\right)
\end{equation}
where the scalar field $W$ on $\mcD$ characterizes the mechanical properties
of the medium. It will be shown how the $1$-form $\Bmu$, map $f$ and
the metric tensor $\Bg$ are used to recover the $4$-velocity vector
field $\Bu$ on $\mcM$. 

\subsection{Equivariance under diffeomorphisms}
Not all scalar fields on $\mcD$ are
admissible candidates for $W$ because not all choices lead to
spacetime covariant variational equations on $\mcM$. The dependence of $\rho$ on
$\gamma^{ab}$ and $\mcF^A_a$ must be such that
$(\SecD)^*(\rho\Bomega)$ is a spacetime coframe independent
$4$-form on $\mcM$. The object $(\SecD)^*\Bomega$ is such a $4$-form on
$\mcM$ by definition so only $(\SecD)^*\rho$ need be considered. 
Geometrically, the local covariance of $(\SecD)^*\rho$ on $\mcM$ may
be formulated in terms of an equivariance property of $\SecD$ acting
on $\rho$ under the group of diffeomorphisms on $\mcM$ since these can
be used to induce local coframe transformations.

Let $\varphi$ be any local diffeomorphism on $\mcM$ with
$\psi\equiv\varphi^{-1}$. Admissible choices for $\rho$ are defined to satisfy
\begin{equation}
\varphi^*\left(\SecD\right)^*\rho = (\DiffSecDabbrev{\varphi})^*\rho
\label{appendix:equivariance}
\end{equation}
where
\begin{equation}
\DiffSecDabbrev{\varphi} \equiv \DiffSecD{\varphi}{\psi}.
\end{equation}
Let $(x^a)$ be the coordinates of $m\in\mcM$ and let $\hat{\rho}$ be the $0$-form
\begin{equation}
\label{appendix:rho_hat}
\hat{\rho}(x^a) \equiv
\rho\left(x^a,f^A(x),\PD_a f^A(x),G^{ab}(x),\lambda(x)\right)
\end{equation}
on $\mcM$ and note that the local coordinate expression of
(\ref{appendix:equivariance}) is
\begin{equation}
\label{appendix:equivariance_local_coordinate}
\begin{split}
\hat{\rho}&\left(\varphi^a(x)\right) = 
\rho\left(\varphi^a(x),(\varphi^*f^A)(x),\PD_a(\varphi^*f^A)(x),
(\psi_*\BG)^{ab}(x),(\varphi^*\lambda)(x)\right)\\
&= \rho\left(\varphi^a(x),(f^A\circ\varphi)(x),
\PD_a(f^A\circ\varphi)(x),(G^{cd}
\PD_c\psi^a\PD_d\psi^b\circ\varphi)(x),(\lambda\circ\varphi)(x)\right)
\end{split}
\end{equation}

Equation (\ref{appendix:equivariance}) may be used to derive
an identity analogous to (\ref{eq:component_equivariance}) at
$m_0\in\mcM$ by a restriction to
diffeomorphisms for which $m_0$ is a fixed point.
At $m=m_0$, where $\varphi^a(x_0)=x^a_0$, using
(\ref{appendix:rho_hat}) and
(\ref{appendix:equivariance_local_coordinate}) it follows that
\begin{equation}
\label{appendix:fixed_point_equivariance}
\rho\left(x^a_0,f^A_0,(\PD_a f^A)_0,G^{ab}_0,\lambda_0\right) =
\rho\left(x^a_0,f^A_0,(S^{-1})^b_{a0} (\PD_b f^A)_0,S^a_{c0} S^b_{d0}
G^{cd}_0,\lambda_0\right)
\end{equation}
where $S^a_{b0}=(\PD_b\psi^a\circ\varphi)(x_0)$,
$f^A_0=f^A(x_0)$, $(\PD_a f^A)_0 = \PD_a f^A(x_0)$,
$G^{ab}_0=G^{ab}(x_0)$ and $l_0=l(x_0)$. Hence, by choosing $\varphi$
appropriately (\ref{appendix:fixed_point_equivariance}) holds at all
points in $\mcM$ for any $S^a_b$ and,
following the same arguments given earlier, may
be used to reduce the dependence of $\rho$:   
\begin{equation}
\rho\left(x^a,f^A(x),\PD_a f^A(x),G^{ab}(x),\lambda(x)\right) =
\bar{\rho}\left(x^a,f^A(x), G^{ab}\PD_a f^A\PD_b f^B(x),\lambda(x)\right)
\end{equation}
for some function $\bar{\rho}$.

An identity analogous to (\ref{eq:derivative_component_equivariance})
may be obtained directly from (\ref{appendix:equivariance}) by using a
$1$-parameter family of local diffeomorphisms $\varphi_{\varepsilon}$
generated by the vector field $\BY$ on $\mcU\subset\mcM$: 
\begin{gather}
\varphi_0(m)=m\quad\forall\, m\in\mcU\subset\mcM,\\
\BY=\frac{d\varphi^a_{\varepsilon}}{d\varepsilon}(x)\biggm|_{\varepsilon=0}
\frac{\PD}{\PD x^a}.
\end{gather}
Using (\ref{appendix:equivariance}) it can be seen that  
\begin{equation}
\frac{d}{d\varepsilon}\varphi_{\varepsilon}^*\left(\SecD\right)^*\rho
\biggm|_{\varepsilon=0} 
=
\frac{d}{d\varepsilon}(\DiffSecDabbrev{\varphi_{\varepsilon}})^*\rho
\biggm|_{\varepsilon=0}
\end{equation}
where $\psi_{\varepsilon}\equiv\varphi^{-1}_{\varepsilon}$ and so
(\ref{appendix:equivariance_local_coordinate}) leads to
\begin{equation}
\label{appendix:Lie_on_rho_hat_1}
\begin{split}
\Lie{\BY}\hat\rho = Y^a\frac{\PD\rho}{\PD x^a} &+
\Lie{\BY}f^A\frac{\PD\rho}{\PD X^A} +
\PD_a(\Lie{\BY}f^A)\frac{\PD\rho}{\PD \mcF^A_a}\\
&+ (\Lie{\BY}\BG)^{ab}\frac{\PD\rho}{\PD\gamma^{ab}}
+ \Lie{\BY}\lambda\frac{\PD\rho}{\PD l}
\end{split}
\end{equation}
where each partial derivative of $\rho$ is implicitly evaluated on
$\SecD$ and $Y^a=dx^a(\BY)$. Furthermore, (\ref{appendix:rho_hat}) yields
\begin{equation}
\label{appendix:Lie_on_rho_hat_2}
\begin{split}
\Lie{\BY}\hat\rho = Y^a\frac{\PD\rho}{\PD x^a} &+
\Lie{\BY}f^A\frac{\PD\rho}{\PD X^A} +
\Lie{\BY}(\PD_a f^A)\frac{\PD\rho}{\PD \mcF^A_a}\\
&+ \Lie{\BY}(G^{ab})\frac{\PD\rho}{\PD\gamma^{ab}}
+ \Lie{\BY}\lambda\frac{\PD\rho}{\PD l}
\end{split}
\end{equation}
which subtracted from (\ref{appendix:Lie_on_rho_hat_1}) leads to 
\begin{equation}
\PD_aY^b\left(\PD_bf^A\frac{\PD\rho}{\PD\mcF^A_a}-
2G^{ac}\frac{\PD\rho}{\PD\gamma^{cb}}\right)=0
\end{equation}
The previous equation holds for any $Y^a$ and so
\begin{equation}
\PD_bf^A\frac{\PD\rho}{\PD\mcF^A_a} = 2G^{ac}\frac{\PD\rho}{\PD\gamma^{cb}}
\end{equation}
and furthermore, using (\ref{appendix:rho_in_terms_of_W}) it follows that
\begin{equation}
\label{appendix:equivariance_condition_on_W}
\PD_bf^A\frac{\PD W}{\PD\mcF^A_a} = 2G^{ac}\frac{\PD W}{\PD\gamma^{cb}}
\end{equation}
on the section $\SecD$ of $\mcD$.
\subsection{The stress-energy tensor and normalization constraint}
The action functional (\ref{appendix:action_functional}) is varied
with respect to $\BG$ via variations of the section $\SecD$ to obtain
the stress-energy tensor $\BT$ and is extremelized with respect to $f$
and $\lambda$ via $\SecD$ to obtain the equation of motion for $f$ and
the normalization constraint $\Bg(\Bu,\Bu)=-1$.

Let $\Gamma_\varepsilon = \SecDGvar$ be the $1$-parameter family of
sections given by
\begin{equation}
\begin{split}
\SecDGvar : \mcM &\rightarrow \mcD\\
x^a &\mapsto (x^a,f^A(x),\PD_a f^A(x),G^{ab}_\varepsilon(x),\lambda(x))
\end{split}
\end{equation}
where $\BG_\varepsilon$ is a $1$-parameter family of contravariant metric tensors on
$\mcM$ and $\BG_0=\BG$. The covariant stress-energy tensor $\BT$ on $\mcM$, dual
to the contravariant stress-energy tensor $\Bt$, of the medium is defined by
\begin{equation}
\frac{d}{d\varepsilon}S[\Gamma_\varepsilon]\bigm|_{\varepsilon=0} =
-\frac{1}{2}\int_\mcM \delta G^{ab}T_{ab}\star 1 
\end{equation}
where $\star$ is the Hodge map associated with the volume
$4$-form $\star 1 =\Gamma_0^*\Bomega$ on $\mcM$,
$T_{ab}\equiv\BT(\PD_a,\PD_b)$, $\Bomega$ is given in (\ref{appendix:omega_form}) and
\begin{equation}
\frac{d}{d\varepsilon}\Gamma_\varepsilon\biggm|_{\varepsilon=0} =
\delta G^{ab} \frac{\PD}{\PD\gamma^{ab}}
\end{equation}
with
\begin{equation}
\delta G^{ab} = \frac{d G^{ab}_\varepsilon}{d\varepsilon}\biggm|_{\varepsilon=0}
\end{equation}
is the variational vector field associated with $\SecDGvar$. Using
\begin{equation}
\delta(\text{det}[\gamma^{ab}]) =
\text{det}[\gamma^{ab}]\gamma_{cd}\delta\gamma^{cd},
\end{equation}
where $\gamma^{ab}\gamma_{bc}=\delta^a_c$ (the
Kronecker delta), leads to
\begin{equation}
\label{appendix:stress_energy_tensor}
T_{ab}= \Gamma^*_0\left(-2\frac{\partial\rho}{\partial \gamma^{ab}} + \rho \gamma_{ab}\right).
\end{equation}
Although $\BT$ has a similar structure to the
stress-energy tensor employed in
\cite{Sch} recall that here $f$ has rank $4$ and $\Gamma^*_0W$ involves
both space and time derivatives of $f$.

Now let $\Gamma_\varepsilon = \SecDlambdavar$ be the $1$-parameter family of
sections given by
\begin{equation}
\begin{split}
\SecDlambdavar : \mcM &\rightarrow \mcD\\
x^a &\mapsto (x^a,f^A(x),\PD_a f^A(x),G^{ab}(x),\lambda_\varepsilon(x))
\end{split}
\end{equation}
where $\lambda_\varepsilon$ is a $1$-parameter family of scalar fields on
$\mcM$ and $\lambda_0=\lambda$. Demanding that
\begin{equation}
\frac{d}{d\varepsilon}S[\Gamma_\varepsilon]\biggm|_{\varepsilon=0} = 0
\end{equation}
for all $\lambda$ variations
\begin{gather}
\frac{d}{d\varepsilon}\Gamma_\varepsilon\biggm|_{\varepsilon=0} =
\delta \lambda \frac{\PD}{\PD l}
\end{gather}
with
\begin{gather}
\delta \lambda = \frac{d \lambda_\varepsilon}{d\varepsilon}\biggm|_{\varepsilon=0}
\end{gather}
leads to
\begin{equation}
\label{appendix:mu_normalization}
\mu_A(f)\mu_B(f)df^A\wedge\star df^B = -\star 1
\end{equation}
where $d$ is the exterior derivative. The vector field $\Bu$
is constructed from the $1$-form $f^*\Bmu = \mu_A(f)df^A$ and the metric $\Bg$ as
\begin{equation}
\Balpha(\Bu) = \BG(f^*\Bmu,\Balpha) 
\end{equation}
where $\Balpha$ is any $1$-form on $\mcM$ and, using equation
(\ref{appendix:mu_normalization}), $\Bu$ satisfies the normalization condition
\begin{equation}
\Bg(\Bu,\Bu) = -1.
\end{equation}
Equation (\ref{appendix:rho_in_terms_of_W}) and
(\ref{appendix:mu_normalization}) are used to write
(\ref{appendix:stress_energy_tensor}) in the form
\begin{equation}
\begin{split}
\label{appendix:stress_energy_tensor1}
T_{ab} &= \Gamma^*_0\left(-\gamma_{ac}\mcF^A_b\frac{\partial W}{\partial \mcF^A_c}
+ W\gamma_{ab} + l\mu_A(X)\mu_B(X)\mcF^A_a\mcF^B_b\right)\\
&= \Gamma^*_0\left(-\gamma_{ac}\mcF^A_b\frac{\partial W}{\partial \mcF^A_c}
+ W\gamma_{ab}\right) + \lambda u_a u_b
\end{split}
\end{equation}
where $u_a$ are the coordinate components of the $1$-form dual
to the vector field $\Bu$. The equations of motion for $f$ can be
obtained in a similar fashion, or derived by setting the divergence of $\BT$ to zero. 


We refer to (\ref{appendix:stress_energy_tensor1}) as the
stress-energy tensor for a {\it simple relativistic hyper-elastic
medium}. In this article the words {\it elastic} and {\it
hyper-elastic} are used purely by analogy with concepts in
classical (non-relativistic) Continuum Mechanics.
Since $\kappa_*$ involves time derivatives this is clearly
an abuse of terminology. 

The derivation of the form of
the hyper-elastic stress-energy tensor
(\ref{appendix:stress_energy_tensor1}) based on
(\ref{appendix:equivariance}) is often referred to as
compatibility with the principle of local covariance on $\mcM$. The subtlety
here is that $(\SecD)^*(\rho \Bomega)$ must be a $4-$form on $\mathcal M$
when constructed out of tensors on the body-time complex
${\mathcal N}$ and the metric tensor $\bf g$ on $\mathcal M$. This
covariance is the general spacetime analogue of classical material objectivity
(for an account of the classical theory see \cite{NLFT}).

In general the stress-energy tensor may not arise from an action
of the form (\ref{appendix:action_functional}) but still may be
written in the form (\ref{eq:elastic})
for some symmetric degree $2$ tensor ${\bf f}$ and where
$n=f(x)=\kappa^{-1}(x)$.

\end{document}